\documentclass[10pt, sigconf]{acmart}

\pdfoutput=1
\usepackage{epsfig,endnotes, enumerate}
\usepackage[normalem]{ulem}
\usepackage{units}
\usepackage[TABBOTCAP]{subfigure}
\usepackage{tabularx}
\usepackage{graphicx} 
\usepackage{color}
\usepackage{xspace}
\usepackage{thumbpdf}
\usepackage{listings}
\usepackage{verbatim}
\usepackage{hyperref}
\usepackage{booktabs}
\usepackage{placeins}
\usepackage{balance}
\usepackage[small,compact]{titlesec}

\usepackage{outlines,url,grffile,enumerate,appendix}
\usepackage{breakurl} 
\usepackage{paralist}
\usepackage{amsmath}
\usepackage{xcolor}
\usepackage{caption}
\usepackage{microtype}
\usepackage{multirow}
\usepackage{framed}
\usepackage{algpseudocode}
\usepackage{longtable}
\usepackage{graphicx}
\usepackage{lipsum}

\renewcommand\footnotetextcopyrightpermission[1]{} 
\setcopyright{none}
\usepackage{enumitem}
\setitemize{itemsep=1pt,topsep=1pt,parsep=1pt,partopsep=1pt}

\newcommand{\bi}{\begin{itemize}}
\newcommand{\ei}{\end{itemize}}

\newcommand{\eg}{{\it e.g.,}\xspace}
\newcommand{\ie}{{\it i.e.,}\xspace}
\newcommand\eat[1]{}
\newcommand\paragraphb[1]{\noindent{\bf #1}}
\newcommand\paragraphi[1]{\noindent\emph{#1}}

\newcommand{\allnotes}[1]{}

\renewcommand{\allnotes}[1]{\textit{#1}}

\newcommand{\fixme}[1]{\allnotes{\bf\textcolor{red}{[#1]}}}
\newcommand{\crchange}[1]{#1}
\newcommand{\notinarxiv}[1]{}

\newcommand{\irn}{IRN\xspace}

\begin{document}
\title{Revisiting Network Support for RDMA}
\subtitle{Extended version of the SIGCOMM 2018 paper}

\author{\large \rm Radhika Mittal$^1$, \rm Alexander Shpiner$^3$,  \rm Aurojit Panda$^4$,  \rm Eitan Zahavi$^3$, \and \rm Arvind Krishnamurthy$^5$, \rm Sylvia Ratnasamy$^1$,  \rm Scott Shenker$^{1,2}$ \and \normalsize \rm $^1$\emph{UC Berkeley}, \rm $^2$\emph{ICSI},  \rm $^3$\emph{Mellanox Technologies}, \rm $^4$\emph{NYU}, \rm $^5$\emph{Univ. of Washington}}
\renewcommand{\shortauthors}{R. Mittal et al.}
\renewcommand{\authors}{Radhika Mittal, Alexander Shpiner, Aurojit Panda, Eitan Zahavi, Arvind Krishnamurthy, Sylvia Ratnasamy, Scott Shenker}




        
\date{}

\begin{abstract}
The advent of RoCE (RDMA over Converged Ethernet)
has led to a significant increase in the use of RDMA in datacenter networks. 
To achieve good performance, RoCE requires a lossless network which is in turn achieved by enabling Priority Flow Control (PFC)  within the network.
However, PFC brings with it a host of problems such as  head-of-the-line blocking, congestion spreading, and occasional deadlocks. 
Rather than seek to fix these issues, we instead ask: \emph{is PFC fundamentally required to support RDMA over Ethernet?} 

We show that the need for PFC is an artifact of current RoCE NIC designs rather than a fundamental requirement. We propose an \emph{improved RoCE NIC} (IRN) design that makes a few simple changes to the RoCE NIC for better handling of packet losses. We show that IRN (without PFC) outperforms RoCE (with PFC) by 6-83\% for typical network scenarios. 
Thus not only does IRN eliminate the need for PFC, it \emph{improves} performance in the process!
We further show that the changes that IRN introduces can be implemented with modest overheads of about 3-10\% to NIC resources.
Based on our results, we argue that research and industry should rethink the current trajectory of network support for RDMA.
\end{abstract}


\maketitle







\sloppy

\section{Introduction}


Datacenter networks offer higher bandwidth and lower latency than traditional wide-area networks. However, traditional endhost networking stacks, with their high latencies and substantial CPU overhead, have limited the extent to which applications can make use of these characteristics. As a result, several large datacenters have recently adopted RDMA, which bypasses the traditional networking stacks in favor of direct memory accesses. 

RDMA over Converged Ethernet (RoCE) has emerged as the canonical method for deploying RDMA in Ethernet-based datacenters \cite{dcqcn, microsoftpfc}. The centerpiece of RoCE is a NIC that (i) provides mechanisms for accessing host memory without CPU involvement and (ii) supports very basic network transport functionality.  Early experience revealed that RoCE NICs only achieve good end-to-end performance when run over a lossless network, so operators turned to Ethernet's Priority Flow Control (PFC) mechanism to achieve minimal packet loss. The combination of RoCE and PFC has enabled a wave of datacenter RDMA deployments.

However, the current solution is not without problems. In particular, 
PFC adds management complexity and can lead to significant performance problems such as head-of-the-line blocking,
congestion spreading, 
and occasional deadlocks \cite{dcqcn, microsoftpfc, tcpbolt, pfcdeadlocks1, pfcdeadlocks2}. Rather than continue down the current path and address the various problems with PFC, in this paper we take a step back and ask whether it was needed in the first place.  
To be clear, current RoCE NICs require a lossless fabric for good performance. However, the question we raise is: \emph{can the RoCE NIC design be altered so that we no longer need a lossless network fabric?} 

We answer this question in the affirmative, proposing a new design called IRN (for Improved RoCE NIC) that makes two incremental changes to current RoCE NICs (i) more efficient loss recovery, and (ii) basic end-to-end flow control to bound the number of in-flight packets (\S\ref{sec:irnTransportDesign}). 
We show, via extensive simulations on a RoCE simulator obtained from a commercial NIC vendor, that IRN performs better than current RoCE NICs, 
and that IRN does not require PFC to achieve high performance; in fact, IRN often performs better without PFC (\S\ref{sec:losslessness-study}). 
We detail the extensions to the RDMA protocol that IRN requires (\S\ref{sec:irn-details}) and use comparative analysis and FPGA synthesis to evaluate the overhead that IRN introduces in terms of NIC hardware resources (\S\ref{sec:feasibility-evaluation}). Our results suggest that adding IRN functionality to current RoCE NICs would add as little as 3-10\% overhead in resource consumption, with no deterioration in message rates.


A natural question that arises is how IRN compares to iWARP? iWARP~\cite{iwarp} long ago proposed a similar philosophy as IRN: handling packet losses efficiently in the NIC rather than making the network lossless. What we show is that iWARP's failing was in its design choices. The differences between iWARP and IRN designs stem from their starting points: iWARP aimed for full generality which led them to put the full TCP/IP stack on the NIC, requiring multiple layers of translation between RDMA abstractions and traditional TCP bytestream abstractions. As a result, iWARP NICs are typically far more complex than RoCE ones, with higher cost and lower performance (\S\ref{sec:background}). In contrast, IRN starts with the much simpler design of RoCE and asks what minimal features can be added to eliminate the need for PFC. 

More generally: while the merits of iWARP vs. RoCE has been a long-running debate in industry, there is no conclusive or rigorous evaluation that compares the two architectures. Instead, RoCE has emerged as the de-facto winner in the marketplace, and brought with it the implicit (and still lingering) assumption that a lossless fabric is necessary to achieve  RoCE's high performance. Our results are the first to rigorously show that, counter to what market adoption might suggest, iWARP in fact had the right architectural philosophy, although a needlessly complex design approach.

Hence, one might view IRN and our results in one of two ways: (i) a new design for RoCE NICs which, at the cost of a few incremental modifications, eliminates the need for PFC and leads to better performance, or, (ii) a new incarnation of the iWARP philosophy which is simpler in implementation and faster in performance.

\eat{As we describe later, we find that with relatively straightforward changes to the RoCE NIC design, well within the limits of feasibility, one can fully deploy RDMA over commodity Ethernet without enabling PFC. Somewhat surprisingly, when running just the minimally featured \fixme{Alex: minimal == without congestion control ?} version of this \emph{improved RoCE NIC} design (henceforth referred to as \irn), PFC is counterproductive, leading to significantly worse performance. When \irn design is paired with advanced congestion control, PFC has little positive impact on performance (and, under some circumstances, sizable negative impact). Given that PFC can lead to many other problems, and does not significantly improve performance when paired with \irn design, we conclude -- contrary to the prevailing wisdom -- that PFC is not fundamentally required for deploying RDMA over Ethernet.
\fixme{we are highlighting our contributions below}}

\section{Background}
\label{sec:background}

We begin with reviewing some relevant background.

\subsection{Infiniband RDMA and RoCE}
RDMA has long been used by the HPC community in special-purpose Infiniband clusters that use credit-based flow control to make the network lossless~\cite{ibaspec}. Because packet drops are rare in such clusters, the RDMA Infiniband transport (as implemented on the NIC) was not designed to efficiently recover from packet losses. When the receiver receives an out-of-order packet, it simply discards it and sends a negative acknowledgement (NACK) to the sender. When the sender sees a NACK, it retransmits all packets that were sent after the last acknowledged packet (i.e., it performs a go-back-N retransmission).  

To take advantage of the widespread use of Ethernet in datacenters, RoCE~\cite{ibaspecroce,ibaspecrocev2} was introduced to enable the use of RDMA over Ethernet.\footnote{We use the term RoCE for both RoCE~\cite{ibaspecroce} and its successor RoCEv2~\cite{ibaspecrocev2} that enables running RDMA, not just over Ethernet, but also over IP-routed networks.} RoCE adopted the same Infiniband transport design (including go-back-N loss recovery), and the network was made lossless using PFC. 

\subsection{Priority Flow Control}
\label{sec:pfc-issues}
Priority Flow Control (PFC)~\cite{pfc} is Ethernet's flow control mechanism, in which a switch sends a pause (or X-OFF) frame to the upstream entity (a switch or a NIC), when the queue exceeds a certain configured threshold. When the queue drains below this threshold, an X-ON frame is sent to resume transmission. When configured correctly, PFC makes the network lossless (as long as all network elements remain functioning). 
However, this coarse reaction to congestion is agnostic to \emph{which} flows are causing it and this results in various performance issues that have been documented in numerous papers in recent years~\cite{dcqcn, microsoftpfc, tcpbolt, pfcdeadlocks1, pfcdeadlocks2}. These issues range from mild (\eg unfairness and head-of-line blocking) to severe, such as ``pause spreading'' as highlighted in \cite{microsoftpfc}
and even network deadlocks~\cite{tcpbolt, pfcdeadlocks1, pfcdeadlocks2}. In an attempt to mitigate these issues, congestion control mechanisms have been proposed for RoCE (\eg DCQCN~\cite{dcqcn} and Timely~\cite{timely}) which reduce the sending rate on detecting congestion, but are not enough to eradicate the need for PFC. Hence, there is now a broad agreement that PFC makes networks harder to understand and manage, and can lead to myriad performance problems that need to be dealt with.

\subsection{iWARP vs RoCE}
iWARP~\cite{iwarp} was designed to support RDMA over a fully general (i.e., not loss-free)  network. iWARP implements the entire TCP stack in hardware along with multiple other layers that it needs to translate TCP's byte stream semantics to RDMA segments. 
Early in our work, we engaged with multiple NIC vendors and datacenter operators in an attempt to understand why 
iWARP was not more broadly adopted (since we believed the basic architectural premise underlying iWARP was correct). 
The consistent response we heard was that iWARP is significantly more complex and expensive than RoCE, with inferior performance~\cite{iwarp-vs-roce}. 

We also looked for empirical datapoints to validate or refute these claims.
We ran RDMA Write benchmarks on two machines connected to one another, using Chelsio T-580-CR 40Gbps iWARP NICs on both machines for one set of experiments, and Mellanox MCX416A-BCAT 56Gbps RoCE NICs (with link speed set to 40Gbps) for another. 
Both NICs had similar specifications, and at the time of purchase, the iWARP NIC cost \$760, while the RoCE NIC cost \$420.
Raw NIC performance values for 64 bytes batched Writes on a single queue-pair are reported in Table~\ref{tab:iwarp-vs-roce-nic}. We find that iWARP has 3$\times$ higher latency and 4$\times$ lower throughput than RoCE.

\begin{table}[t]
\centering
\scriptsize
\renewcommand{\arraystretch}{1.25}
\newcolumntype{P}[1]{>{\centering\arraybackslash}p{#1}}
\setlength{\tabcolsep}{3.2pt}
\begin{tabular}[b]{|P{4cm}|P{1.5cm}|P{1.5cm}|}
\hline
\textbf{NIC} & \textbf{Throughput} & \textbf{Latency} \\
\hline
Chelsio T-580-CR (iWARP) & 3.24 Mpps & 2.89 $\mu$s \\
Mellanox MCX416A-BCAT (RoCE) &  14.7 Mpps & 0.94 $\mu$s \\
\hline
\end{tabular}
\caption{An iWARP and a RoCE NIC's raw performance for 64B RDMA Writes on a single queue-pair.}
\vspace{-15pt}
\label{tab:iwarp-vs-roce-nic}
\end{table}



These price and performance differences could be attributed to many factors other than transport design complexity (such as differences in profit margins, supported features and engineering effort)  and hence should be viewed as anecdotal evidence as best.
Nonetheless, they show that our conjecture (in favor of implementing loss recovery at the endhost NIC) was certainly not obvious based on current iWARP NICs. 

\eat{ 

 only anecdotal evidence in support of the claim since the differences that we observe c
Nonetheless, at least anecdotally they certainly do nothing to refute the claims we heard from vendors and operators that iWARP as we learned from conversations with multiple NIC vendors and large customers, this inefficiency of iWARP, combined with the implicit assumption that losslessness results in better performance, made RoCE with PFC the current canonical choice for deploying RDMA over Ethernet.

\vspace{10pt}
} 

Our primary contribution is to show that iWARP, somewhat surprisingly, did in fact have the right philosophy: explicitly handling packet losses in the NIC leads to better performance than having a lossless network. However, efficiently handling packet loss does not require implementing the entire TCP stack in hardware as iWARP did. Instead, we identify the incremental changes to be made to current RoCE NICs, leading to a design which (i) does not require PFC yet achieves better network-wide performance than both RoCE and iWARP (\S\ref{sec:losslessness-study}), and (ii) is much closer to RoCE's implementation with respect to both NIC performance and complexity (\S\ref{sec:feasibility-evaluation}) and is thus significantly less complex than iWARP.

\section{IRN Design}
\label{sec:irnTransportDesign}

We begin with describing the transport logic for IRN. For simplicity, we present it as a general design independent of the specific RDMA operation types. We go into the details of handling specific RDMA operations with IRN later in \S\ref{sec:irn-details}. 

\crchange{Changes to the RoCE transport design may introduce overheads in the form of new hardware logic or additional per-flow state. With the goal of keeping such overheads as small as possible, IRN strives to make \emph{minimal} changes to the RoCE NIC design in order to eliminate its PFC requirement, 
as opposed to squeezing out the best possible performance with a more sophisticated design (we evaluate the small overhead introduced by IRN later in \S\ref{sec:feasibility-evaluation}).}

IRN, therefore, makes two key changes to current RoCE NICs, as described in the following subsections: (1) improving the loss recovery mechanism, and (2) basic end-to-end flow control (termed BDP-FC) which bounds the number of in-flight packets by the bandwidth-delay product of the network. \crchange{We justify these changes by empirically evaluating their significance, and exploring some alternative design choices later in \S\ref{sec:factor-analysis}.} Note that these changes are orthogonal to the use of explicit congestion control mechanisms (such as DCQCN~\cite{dcqcn} and Timely~\cite{timely}) that, as with current RoCE NICs, can be \crchange{\emph{optionally}} enabled with IRN.


\subsection{IRN's Loss Recovery Mechanism}
\label{sec:irnlossrecovery}

As discussed in \S\ref{sec:background}, current RoCE NICs use a go-back-N loss recovery scheme. In the absence of PFC, redundant retransmissions caused by go-back-N loss recovery result in significant performance penalties (as evaluated in \S\ref{sec:losslessness-study}). Therefore, the first change we make with IRN is a more efficient loss recovery, based on selective retransmission (inspired by TCP's loss recovery), where the receiver does not discard out of order packets and the sender selectively retransmits the lost packets, as detailed below. 





Upon every out-of-order packet arrival, an IRN receiver sends a NACK, which carries both the cumulative acknowledgment (indicating its expected sequence number) and the sequence number of the packet that triggered the NACK (as a simplified form of selective acknowledgement or SACK). 

An IRN sender enters loss recovery mode when a NACK is received or when a timeout occurs. It also maintains a bitmap to track which packets have been cumulatively and selectively acknowledged. When in the loss recovery mode, the sender selectively retransmits lost packets as indicated by the bitmap, instead of sending new packets. The first packet that is retransmitted on entering loss recovery corresponds to the cumulative acknowledgement value. Any subsequent packet is considered lost only if another packet with a higher sequence number has been selectively acked. When there are no more lost packets to be retransmitted, the sender continues to transmit new packets (if allowed by BDP-FC). It exits loss recovery when a cumulative acknowledgement greater than the \emph{recovery sequence} is received, where the recovery sequence corresponds to the last regular packet that was sent before the retransmission of a lost packet.

SACKs allow efficient loss recovery only when there are multiple packets in flight. For other cases (\eg for single packet messages), loss recovery gets triggered via timeouts. A high timeout value can increase the tail latency of such short messages. However, keeping the timeout value too small can result in too many spurious retransmissions, affecting the overall results. An IRN sender, therefore, uses a low timeout value of $RTO_{low}$ only when there are a small $N$ number of packets in flight \crchange{(such that spurious retransmissions remains negligibly small)}, and a higher value of $RTO_{high}$ otherwise. We discuss \crchange{how the values of these parameters are set in \S\ref{sec:losslessness-study}, and} how the timeout feature in current RoCE NICs can be easily extended to support this in \S\ref{sec:feasibility-evaluation}. 

\subsection{IRN's BDP-FC Mechanism}
\label{sec:irnbdpfc}

The second change we make with IRN is introducing the notion of a basic end-to-end packet level flow control, called BDP-FC, which bounds the number of outstanding packets in flight for a flow by the bandwidth-delay product (BDP) of the network, as suggested in~\cite{pfabric}. This is a static cap that we compute by dividing the BDP of the longest path in the network (in bytes)  \footnote{As in~\cite{pfabric}, we expect this information to be available in a datacenter setting with known topology and routes. IRN does not require a fully precise BDP computation and over-estimating the BDP value would still provide the required benefits to a large extent without under-utilizing the network.} with the packet MTU set by the RDMA queue-pair (typically 1KB in RoCE NICs). An IRN sender transmits a new packet only if the number of packets in flight (computed as the difference between current packet's sequence number and last acknowledged sequence number) is less than this BDP cap. 

BDP-FC improves the performance by reducing unnecessary queuing in the network. Furthermore, by strictly upper bounding the number of out-of-order packet arrivals, it greatly reduces the amount of state required for tracking packet losses in the NICs (discussed in more details in \S\ref{sec:feasibility-evaluation}).


\vspace{10pt}

\noindent As mentioned before, IRN's loss recovery has been inspired by \eat{conventional } TCP's loss recovery. However, rather than incorporating the entire TCP stack as is done by iWARP NICs, IRN: (1) decouples loss recovery from congestion control and does not incorporate any notion of TCP congestion window control involving slow start, AIMD or advanced fast recovery,
(2) operates directly on RDMA segments instead of using TCP's byte stream abstraction, which not only avoids the complexity introduced by multiple translation layers (as needed in iWARP), but also allows IRN to simplify its selective acknowledgement and loss tracking schemes. We discuss how these changes effect performance towards the end of \S\ref{sec:losslessness-study}. 


\section{Evaluating IRN's Transport Logic}
\label{sec:losslessness-study}

We now confront the central question of this paper: {\em Does RDMA require a lossless network?} If the answer is yes, then we must address the many difficulties of PFC.
If the answer is no, then we can greatly simplify network management by letting go of PFC. 
To answer this question, we evaluate the network-wide performance of IRN's transport logic via extensive simulations. Our results show that IRN performs better than RoCE, without requiring PFC. We test this across a wide variety of experimental scenarios and across different performance metrics. We end this section with a simulation-based comparison of IRN with Resilient RoCE~\cite{roceRocks} and iWARP~\cite{iwarp}.

\subsection{Experimental Settings}
\label{sec:exptSettings}

We begin with describing our experimental settings.

\paragraphb{Simulator:} Our simulator, obtained from a commercial NIC vendor, extends INET/OMNET++~\cite{omnet, inet} to model the Mellanox ConnectX4 RoCE NIC~\cite{mlxcx4}. 
RDMA queue-pairs (QPs) are modelled as UDP applications with either RoCE or \irn transport layer logic, that generate flows (as described later).
We define a flow as a unit of data transfer comprising of one or more messages between the same source-destination pair as in~\cite{timely, dcqcn}.
When the sender QP is ready to transmit data packets, it periodically polls the MAC layer until the link is available for transmission.
The simulator implements DCQCN as implemented in the Mellanox ConnectX-4 ROCE NIC~\cite{roceRocks}, and we add support for a NIC-based Timely implementation. 
All switches in our simulation are input-queued with virtual output ports, that are scheduled using round-robin. The switches can be configured to generate PFC frames by setting appropriate buffer thresholds. 




\eat{The X-OFF PFC frame for infinite duration is triggered by the switch to pause the incoming traffic on a link when the input-queue size hits a pre-determined threshold. As the input-queue drains to below that threshold, an X-ON PFC frame is triggered by the switch to resume the incoming traffic.} 

\paragraphb{Default Case Scenario:} 
For our default case, we simulate a 54-server three-tiered fat-tree topology, connected by a fabric with full bisection-bandwidth constructed from  45 6-port switches organized into 6 pods~\cite{pfabricHotnets}. We consider 40Gbps links, each with a propagation delay of 2$\mu$s,
resulting in a bandwidth-delay product (BDP) of 120KB along the longest (6-hop) path. This corresponds to $\sim$110 MTU-sized packets (assuming typical RDMA MTU of 1KB).

Each end host generates new flows with Poisson inter-arrival times~\cite{pfabric, rc3}. Each flow's destination is picked randomly and size is drawn from a realistic heavy-tailed distribution derived from~\cite{adityaflowsizes}. Most flows are small (50\% of the flows are single packet messages with sizes ranging between 32 bytes-1KB representing small RPCs such as those generated by RDMA based key-value stores~\cite{farm, herd}), and most of the bytes are in large flows (15\% of the flows are between 200KB-3MB, representing background RDMA traffic such as storage). 
The network load is set at 70\% utilization for our default case. We use ECMP for load-balancing~\cite{microsoftpfc}. We vary different aspects from our default scenario (including topology size, workload pattern and link utilization) in \S\ref{sec:robustness}.


\paragraphb{Parameters:} $RTO_{high}$ is set to an estimation of the maximum round trip time with one congested link. We compute this as the sum of the propagation delay on the longest path and the maximum queuing delay a packet would see if the switch buffer on a congested link is completely full. This is approximately 320$\mu$s for our default case. For IRN, we set $RTO_{low}$ to 100$\mu$s (representing the desirable upper-bound on tail latency for short messages) with $N$ set to a small value of 3. 
When using RoCE without PFC, we use a fixed timeout value of $RTO_{high}$. 
We disable timeouts when PFC is enabled to prevent spurious retransmissions. We use buffers sized at twice the BDP of the network (which is 240KB in our default case) for each input port~\cite{pfabric, mckeownbuff}.
The PFC threshold at the switches is set to the buffer size minus a headroom equal to the upstream link's bandwidth-delay product (needed to absorb all packets in flight along the link). This is 220KB for our default case. 
We vary these parameters in \S\ref{sec:robustness} to show that our results are not very sensitive to these specific choices. 
When using RoCE or IRN with Timely or DCQCN, we use the same congestion control parameters as specified in~\cite{timely} and~\cite{dcqcn} respectively. For fair comparison with PFC-based proposals~\cite{dcqcn, tcpbolt}, the flow starts at line-rate for all cases.

\paragraphb{Metrics:} 
We primarily look at three metrics:
 (i) average slowdown, where slowdown for a flow is its completion time divided by the time it would have taken to traverse its path at line rate in an empty network, (ii) average flow completion time (FCT), (iii) 99\%ile or tail FCT. While the average and tail FCTs are dominated by the performance of throughput-sensitive flows, the average slowdown is dominated by the performance of latency-sensitive short flows.

\subsection{Basic Results}
\label{sec:basicResults}

We now present our basic results comparing IRN and RoCE for our default scenario. Unless otherwise specified, IRN is always used without PFC, while RoCE is always used with PFC for the results presented here.

\begin{figure}[t]
\centering
 \renewcommand{\arraystretch}{0.1}
 \footnotesize

\renewcommand{\arraystretch}{1.0}
\newcolumntype{P}[1]{>{\centering\arraybackslash}p{#1}}
\setlength{\tabcolsep}{2pt}
\begin{tabular}{ccc}

\multicolumn{3}{c}{

 \begin{subfigure}
 \centering
\includegraphics[height=0.04\textwidth]{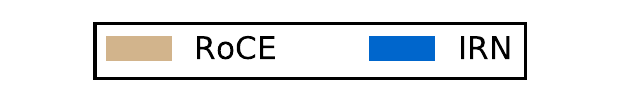}
 \end{subfigure} 

} \vspace{-10pt} \\

 \begin{subfigure}
 \centering
\includegraphics[width=0.125\textwidth,height=0.12\textwidth]{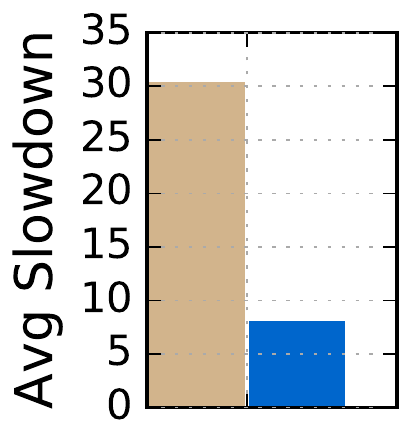}
 \end{subfigure} 
 
 &
 
  \begin{subfigure}
\centering
\includegraphics[width=0.125\textwidth,height=0.12\textwidth]{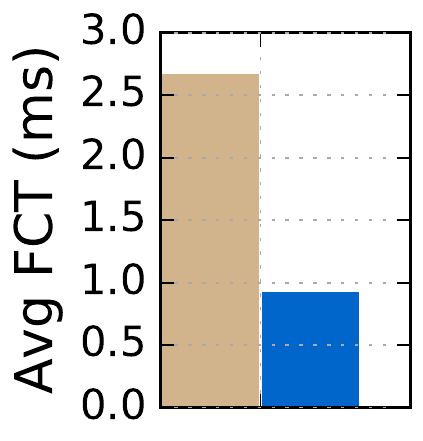}
\end{subfigure}

&

\begin{subfigure}
\centering
\includegraphics[width=0.125\textwidth,height=0.12\textwidth]{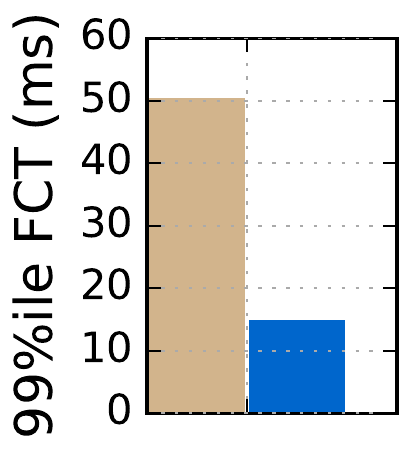}
\end{subfigure}

\end{tabular}

\caption{Comparing IRN and RoCE's performance.
}

\vspace{-10pt}
\label{fig:irn-vs-roce-nocc}
\end{figure}

\begin{figure}[t]
\centering
 \renewcommand{\arraystretch}{0.1}
 \footnotesize

\renewcommand{\arraystretch}{1.0}
\newcolumntype{P}[1]{>{\centering\arraybackslash}p{#1}}
\setlength{\tabcolsep}{2pt}
\begin{tabular}{ccc}

\multicolumn{3}{c}{

 \begin{subfigure}
 \centering
\includegraphics[height=0.04\textwidth]{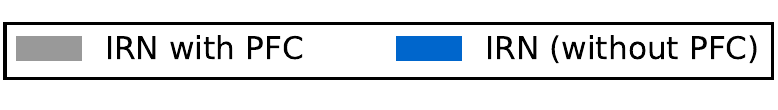}
 \end{subfigure} 

} \vspace{-10pt} \\

 \begin{subfigure}
 \centering
\includegraphics[width=0.125\textwidth,height=0.12\textwidth]{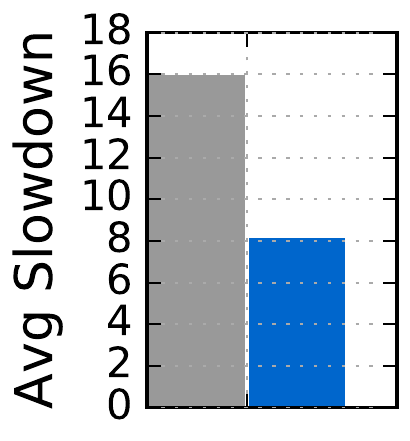}
 \end{subfigure} 
 
 &
 
  \begin{subfigure}
\centering
\includegraphics[width=0.125\textwidth,height=0.12\textwidth]{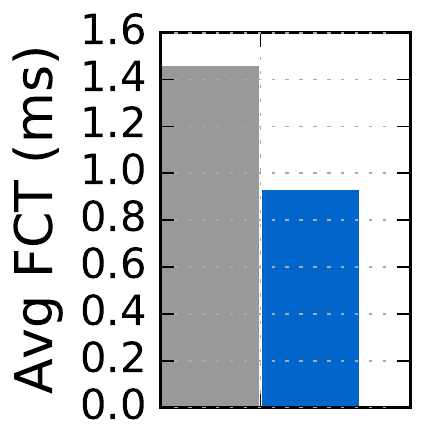}
\end{subfigure}

&

\begin{subfigure}
\centering
\includegraphics[width=0.125\textwidth,height=0.12\textwidth]{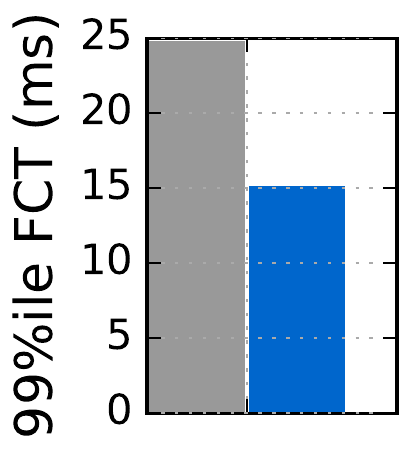}
\end{subfigure}

\end{tabular}

\caption{Impact of enabling PFC with IRN.
}

\vspace{-10pt}
\label{fig:irn-nocc}
\end{figure}

\begin{figure}[t]
\centering
 \renewcommand{\arraystretch}{0.1}
 \footnotesize

\renewcommand{\arraystretch}{1.0}
\newcolumntype{P}[1]{>{\centering\arraybackslash}p{#1}}
\setlength{\tabcolsep}{2pt}
\begin{tabular}{ccc}

\multicolumn{3}{c}{

 \begin{subfigure}
 \centering
\includegraphics[height=0.04\textwidth]{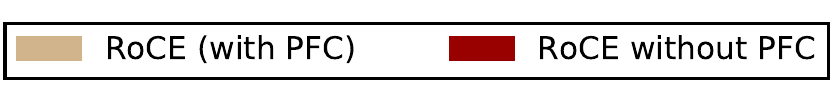}
 \end{subfigure} 

} \vspace{-10pt} \\

 \begin{subfigure}
 \centering
\includegraphics[width=0.125\textwidth,height=0.12\textwidth]{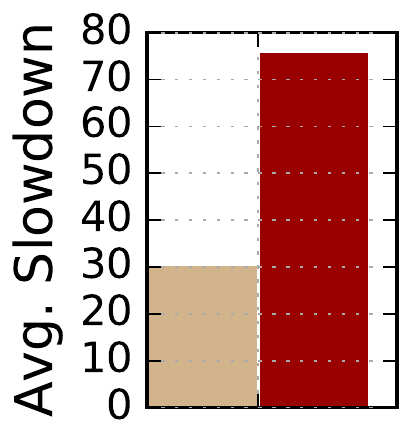}
 \end{subfigure} 
 
 &
 
  \begin{subfigure}
\centering
\includegraphics[width=0.125\textwidth,height=0.12\textwidth]{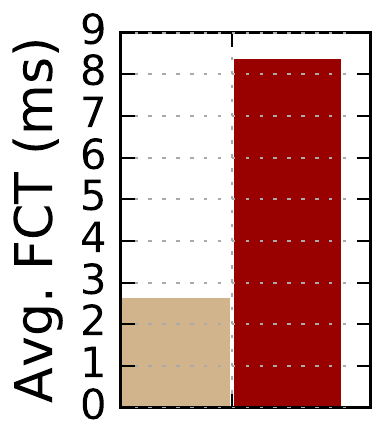}
\end{subfigure}

&

\begin{subfigure}
\centering
\includegraphics[width=0.125\textwidth,height=0.12\textwidth]{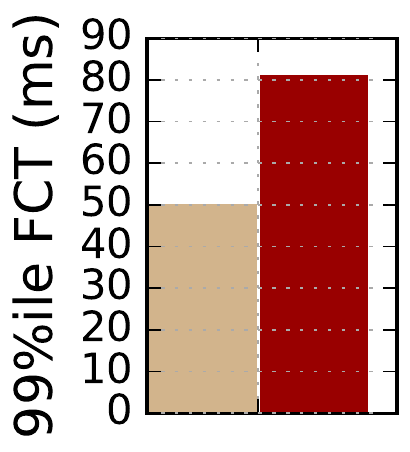}
\end{subfigure}

\end{tabular}

\caption{Impact of disabling PFC with RoCE.
}

\vspace{-10pt}
\label{fig:roce-nocc}
\end{figure}

\subsubsection{IRN performs better than RoCE.}

We begin with comparing IRN's performance with current RoCE NIC's. The results are shown in Figure~\ref{fig:irn-vs-roce-nocc}. IRN's performance is upto 2.8-3.7$\times$ better than RoCE across the three metrics. This is due to the combination of two factors: (i) IRN's BDP-FC mechanism reduces unnecessary queuing and (ii) unlike RoCE, IRN does not experience any congestion spreading issues, since it does not use PFC. (explained in more details below).

\subsubsection{IRN does not require PFC.}

We next study how IRN's performance is impacted by enabling PFC. If enabling PFC with IRN does not improve performance, we can conclude that IRN's loss recovery is sufficient to eliminate the requirement for PFC. However, if enabling PFC with IRN significantly improves performance, we would have to conclude that PFC continues to be important, even with IRN's loss recovery. Figure~\ref{fig:irn-nocc} shows the results of this comparison. Remarkably, we find that not only is PFC not required, but it significantly degrades \irn's performance (increasing the value of each metric by about 1.5-2$\times$). This is because of the head-of-the-line blocking and congestion spreading issues PFC is notorious for: pauses triggered by congestion at one link, cause queue build up and pauses at other upstream entities, creating a cascading effect. Note that, without PFC, IRN experiences significantly high packet drops (8.5\%), which also have a negative impact on performance, since it takes about one round trip time to detect a packet loss and another round trip time to recover from it. However, the negative impact of a packet drop (given efficient loss recovery), is restricted to the flow that faces congestion and does not spread to other flows, as in the case of PFC. While these PFC issues have been observed before~\cite{microsoftpfc, dcqcn, timely}, we believe our work is the first to show that \emph{a well-design loss-recovery mechanism outweighs a lossless network}. 

\subsubsection{RoCE requires PFC.}

Given the above results, the next question one might have is whether RoCE required PFC in the first place? Figure~\ref{fig:roce-nocc} shows the performance of RoCE with and without PFC. We find that the use of PFC helps considerably here. Disabling PFC degrades performance by 1.5-3$\times$ across the three metrics. This is because of the go-back-N loss recovery used by current RoCE NICs, which penalizes performance due to (i) increased congestion caused by redundant retransmissions and (ii) the time and bandwidth wasted by flows in sending these redundant packets.

\begin{figure}[t]
\centering
 \renewcommand{\arraystretch}{0.1}
 \footnotesize

\renewcommand{\arraystretch}{1.0}
\newcolumntype{P}[1]{>{\centering\arraybackslash}p{#1}}
\setlength{\tabcolsep}{2pt}
\begin{tabular}{ccc}

\multicolumn{3}{c}{

 \begin{subfigure}
 \centering
\includegraphics[height=0.04\textwidth]{figures/irn-vs-roce/legend-roce-irn.pdf}
 \end{subfigure} 

} \vspace{-10pt} \\

 \begin{subfigure}
 \centering
\includegraphics[width=0.15\textwidth,height=0.12\textwidth]{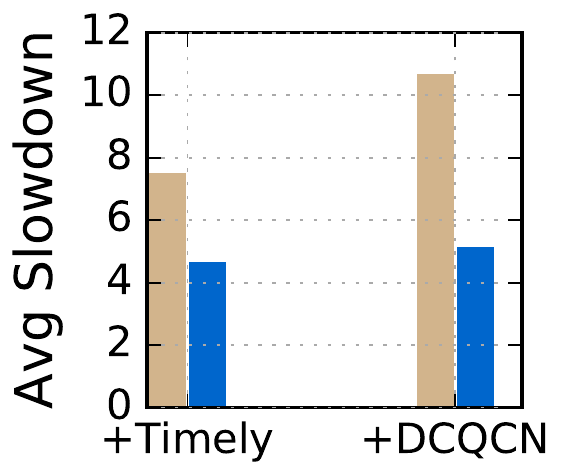}
 \end{subfigure} 
 
 &
 
  \begin{subfigure}
\centering
\includegraphics[width=0.15\textwidth,height=0.12\textwidth]{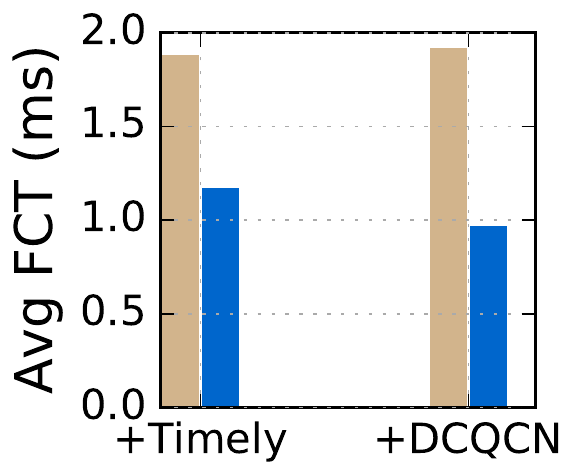}
\end{subfigure}

&

\begin{subfigure}
\centering
\includegraphics[width=0.15\textwidth,height=0.12\textwidth]{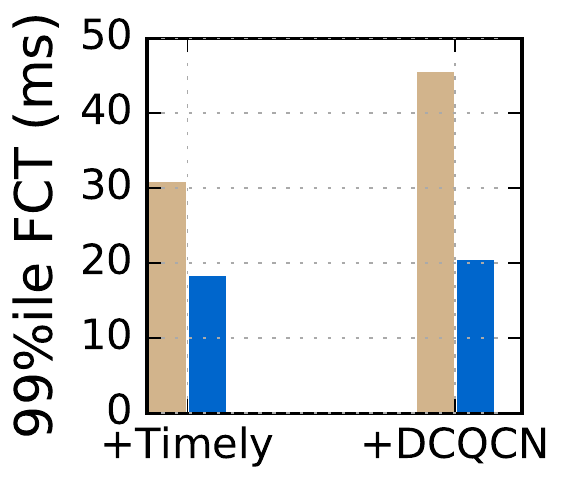}
\end{subfigure}

\end{tabular}

\caption{Comparing IRN and RoCE's performance with explicit congestion control (Timely and DCQCN).
}

\vspace{-10pt}
\label{fig:irn-vs-roce-withcc}
\end{figure}

\begin{figure}[t]
\centering
 \renewcommand{\arraystretch}{0.1}
 \footnotesize

\renewcommand{\arraystretch}{1.0}
\newcolumntype{P}[1]{>{\centering\arraybackslash}p{#1}}
\setlength{\tabcolsep}{2pt}
\begin{tabular}{ccc}

\multicolumn{3}{c}{

 \begin{subfigure}
 \centering
\includegraphics[height=0.04\textwidth]{figures/irn/legend-irn-pfc.pdf}
 \end{subfigure} 

} \vspace{-10pt} \\

 \begin{subfigure}
 \centering
\includegraphics[width=0.15\textwidth,height=0.12\textwidth]{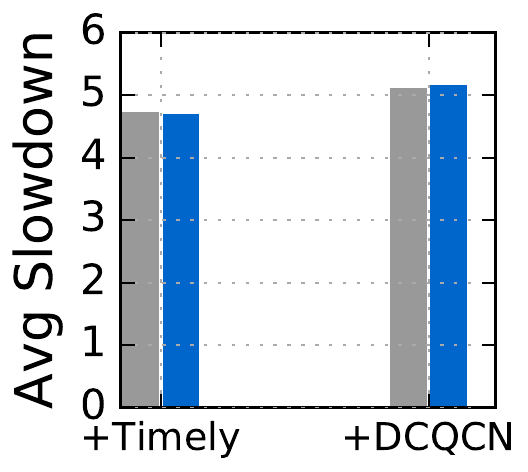}
 \end{subfigure} 
 
 &
 
  \begin{subfigure}
\centering
\includegraphics[width=0.15\textwidth,height=0.12\textwidth]{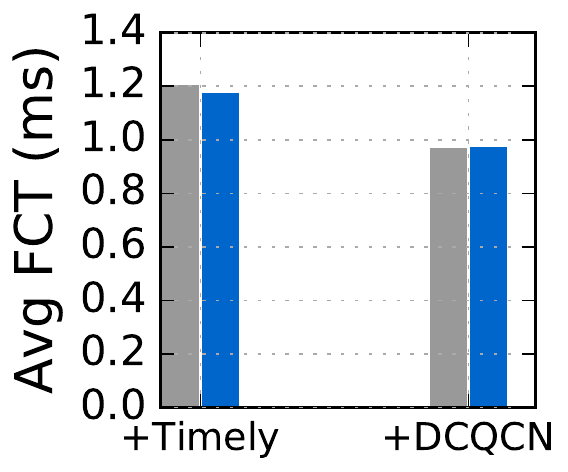}
\end{subfigure}

&

\begin{subfigure}
\centering
\includegraphics[width=0.15\textwidth,height=0.12\textwidth]{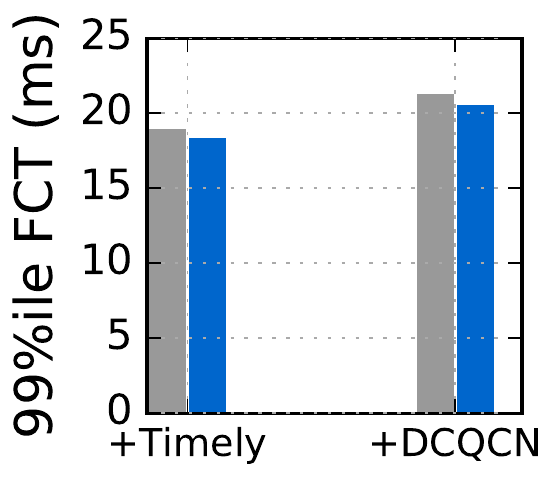}
\end{subfigure}

\end{tabular}

\caption{Impact of enabling PFC with IRN, when explicit congestion control (Timely and DCQCN) is used.
}

\vspace{-10pt}
\label{fig:irn-withcc}
\end{figure}

\begin{figure}[t]
\centering
 \renewcommand{\arraystretch}{0.1}
 \footnotesize

\renewcommand{\arraystretch}{1.0}
\newcolumntype{P}[1]{>{\centering\arraybackslash}p{#1}}
\setlength{\tabcolsep}{2pt}
\begin{tabular}{ccc}

\multicolumn{3}{c}{

 \begin{subfigure}
 \centering
\includegraphics[height=0.04\textwidth]{figures/roce/legend-roce-pfc.pdf}
 \end{subfigure} 

} \vspace{-10pt} \\

 \begin{subfigure}
 \centering
\includegraphics[width=0.15\textwidth,height=0.12\textwidth]{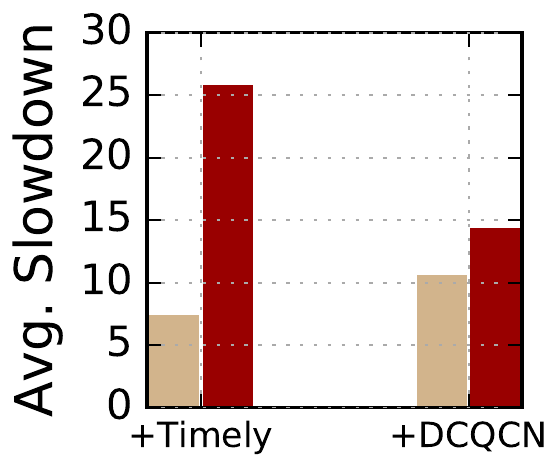}
 \end{subfigure} 
 
 &
 
  \begin{subfigure}
\centering
\includegraphics[width=0.15\textwidth,height=0.12\textwidth]{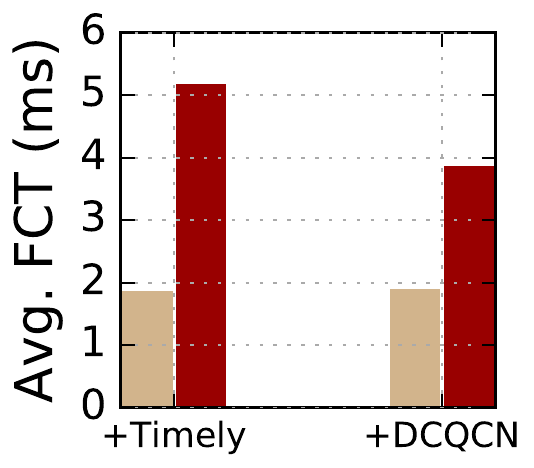}
\end{subfigure}

&

\begin{subfigure}
\centering
\includegraphics[width=0.15\textwidth,height=0.12\textwidth]{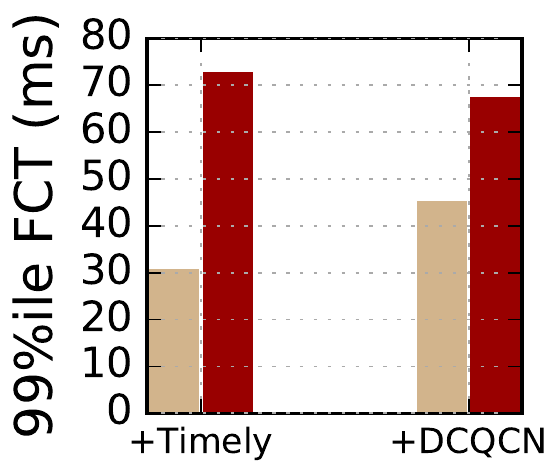}
\end{subfigure}

\end{tabular}

\caption{Impact of disabling PFC with RoCE, when explicit congestion control (Timely and DCQCN) is used.
}

\vspace{-10pt}
\label{fig:roce-withcc}
\end{figure}

\subsubsection{Effect of Explicit Congestion Control.}

The previous comparisons did not use any explicit congestion control. However, as mentioned before, RoCE today is typically deployed in conjunction with some explicit congestion control mechanism such as Timely or DCQCN. We now evaluate whether using such explicit congestion control mechanisms affect the key trends described above.

Figure~\ref{fig:irn-vs-roce-withcc} compares IRN and RoCE's performance when Timely or DCQCN is used. IRN continues to perform better by up to 1.5-2.2$\times$ across the three metrics.

Figure~\ref{fig:irn-withcc} evaluates the impact of enabling PFC with IRN, when Timely or DCQCN is used. We find that, \irn's performance is largely unaffected by PFC, since explicit congestion control mitigates both packet drop rate as well as the number of pause frames generated. The largest performance improvement due to enabling PFC was less than 1\%, while its largest negative impact was about 3.4\%. 

Finally, Figure~\ref{fig:roce-withcc} compares RoCE's performance with and without PFC, when Timely or DCQCN is used.\footnote{\crchange{RoCE + DCQCN without PFC presented in Figure~\ref{fig:roce-withcc} is equivalent to Resilient RoCE~\cite{roceRocks}. We provide a direct comparison of IRN with Resilient RoCE later in this section.}} We find that, unlike IRN, RoCE (with its inefficient go-back-N loss recovery) requires PFC, even when explicit congestion control is used. Enabling PFC improves RoCE's performance by 1.35$\times$ to 3.5$\times$ across the three metrics. 

\vspace{10pt}

\paragraphb{Key Takeaways:} The following are, therefore, the three takeaways from these results: (1) IRN (without PFC) performs better than RoCE (with PFC), (2) IRN does not require PFC, and (3) RoCE requires PFC.

\subsection{Factor Analysis of IRN} 
\label{sec:factor-analysis}

\begin{figure}[t]

\centering
\includegraphics[width=0.45\textwidth]{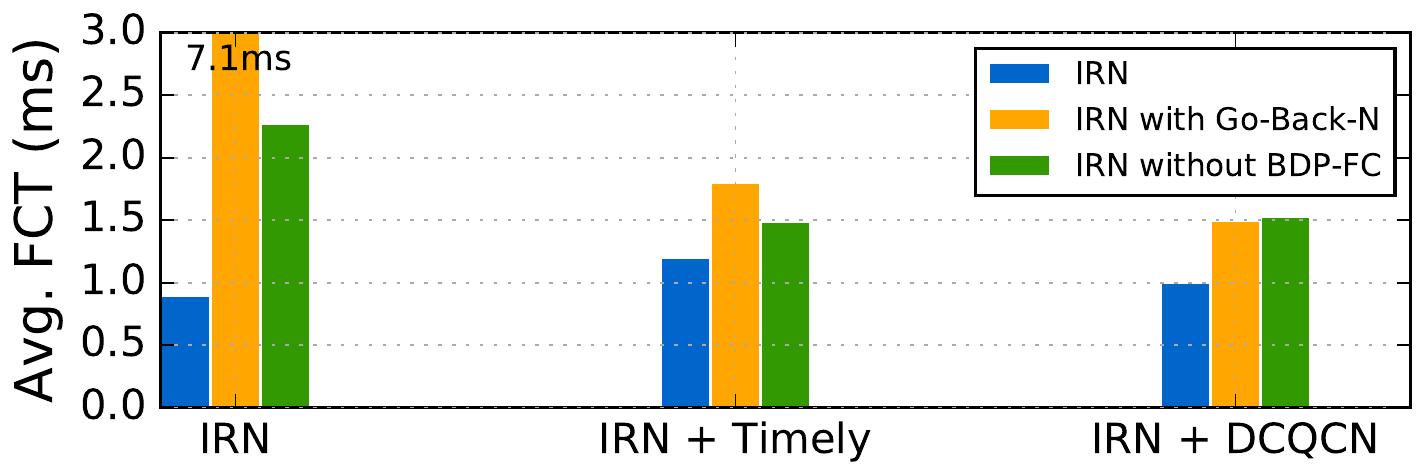}

\caption{The figure shows the effect of doing go-back-N loss recovery and disabling BDP-FC with IRN. The y-axis is capped at 0.003s to better highlight the trends.}
\vspace{-5pt}

\label{fig:factorAnalysis}
\end{figure}

We now perform a factor analaysis of IRN, to individually study the significance of the two key changes IRN makes to RoCE, namely (1) efficient loss recovery and (2) BDP-FC. For this we compare IRN's performance (as evaluated in \S\ref{sec:basicResults}) with two different variations that highlight the significance of each change: (1) enabling go-back-N loss recovery instead of using SACKs, and (2) disabling BDP-FC. Figure~\ref{fig:factorAnalysis} shows the resulting average FCTs (we saw similar trends with other metrics). We discuss these results in greater details below. 


\paragraphb{Need for Efficient Loss Recovery:} The first two bars in Figure~\ref{fig:factorAnalysis} compare the average FCT of default SACK-based IRN and IRN with go-back-N respectively. We find that the latter results in significantly worse performance. This is because of the bandwidth wasted by go-back-N due to redundant retransmissions, as described before.  

Before converging to IRN's current loss recovery mechanism, we experimented with alternative designs. In particular we explored the following questions: 

\paragraphi{(1) Can go-back-N be made more efficient?} 
Go-back-N does have the advantage of simplicity over selective retransmission, since it allows the receiver to simply discard out-of-order packets. We, therefore, tried to explore whether we can mitigate the negative effects of go-back-N.
We found that explicitly backing off on losses improved go-back-N performance for Timely (though, not for DCQCN). Nonetheless, SACK-based loss recovery continued to perform significantly better across different scenarios (with the difference in average FCT for Timely ranging from 20\%-50\%). 

\paragraphi{(2) Do we need SACKs?} We tried a selective retransmit scheme without SACKs (where the sender does not maintain a bitmap to track selective acknowledgements). This performed better than go-back-N. However, it fared poorly when there were multiple losses in a window, requiring multiple round-trips to recover from them. The corresponding degradation in average FCT ranged from $<$1\% up to 75\% across different scenarios when compared to SACK-based IRN.

\paragraphi{(3) \crchange{Can the timeout value be computed dynamically?}} \crchange{As described in \S\ref{sec:irnTransportDesign}, IRN uses two static (low and high) timeout values to allow faster recovery for short messages, while avoiding spurious retransmissions for large ones. We also experimented with an alternative approach of using dynamically computed timeout values (as with TCP), which not only complicated the design, but did not help since these effects were then be dominated by the initial timeout value. }


\begin{figure*}[!ht]

\centering
 \renewcommand{\arraystretch}{0.1}
 \setlength{\tabcolsep}{0.5pt}
 \footnotesize

\begin{tabular}{ccc}

 \begin{subfigure}
 \centering
\includegraphics[width=0.33\textwidth]{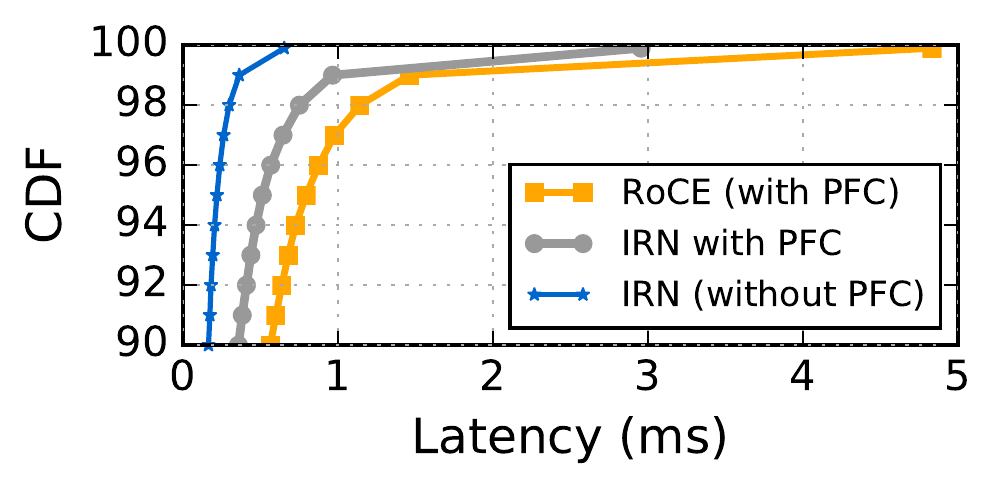}
 \end{subfigure}

&

\begin{subfigure}
\centering
\includegraphics[width=0.33\textwidth]{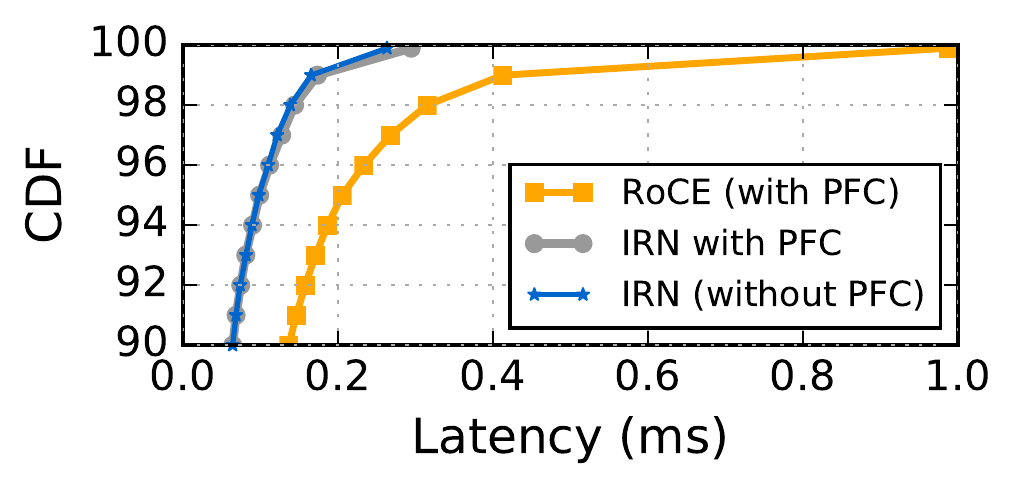}
\end{subfigure}

&

\begin{subfigure}
\centering
\includegraphics[width=0.33\textwidth]{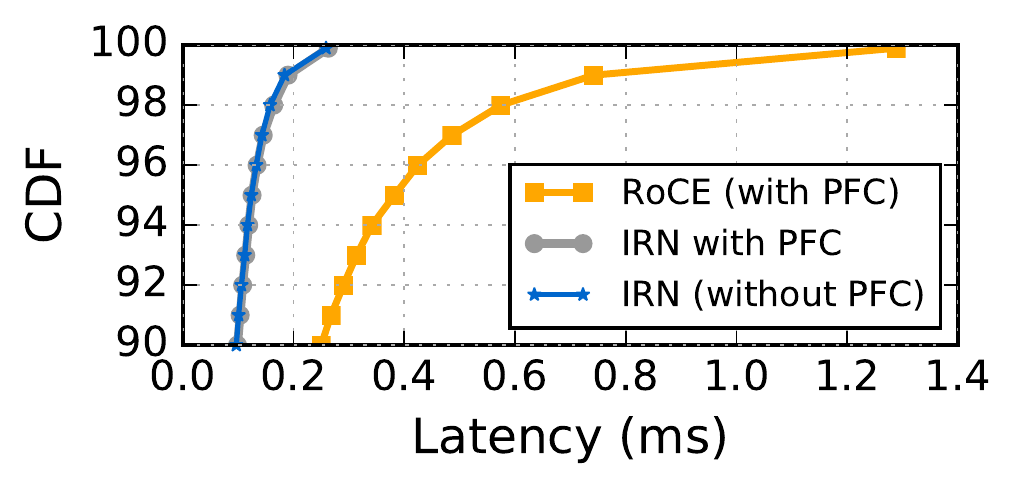}
\end{subfigure}
\\

 (a) No CC & (b) Timely & (c) DCQCN

\end{tabular}

\caption{The figures compare the tail latency for single-packet messages for IRN, IRN with PFC, and RoCE (with PFC), across different congestion control algorithms.}
\vspace{-10pt}

\label{fig:rpc-tail-latency}
\end{figure*}

\paragraphb{Significance of BDP-FC:}  
 The first and the third bars in Figure~\ref{fig:factorAnalysis} compare the average FCT of IRN with and without BDP-FC respectively. We find that BDP-FC significantly improves performance by reducing unnecessary queuing. Furthermore, it prevents a flow that is recovering from a packet loss from sending additional new packets and increasing congestion, until the loss has been recovered. 
 
\paragraphb{Efficient Loss Recovery vs BDP-FC:} Comparing the second and third bars in Figure~\ref{fig:factorAnalysis} shows that the performance of IRN with go-back-N loss recovery is generally worse than the performance of IRN without BDP-FC. This indicates that of the two changes IRN makes, efficient loss recovery helps performance more than BDP-FC. 
 

\subsection{Robustness of Basic Results}
\label{sec:robustness}

We now evaluate the robustness of the basic results from \S\ref{sec:basicResults} across different scenarios and performance metrics.

\subsubsection{Varying Experimental Scenario.}

We evaluate the robustness of our results, as the experimental scenario is varied from our default case. In particular, we run experiments with (i) link utilization levels varied between 30\%-90\%, (ii) link bandwidths varied from the default of 40Gbps to 10Gbps and 100Gbps, (iii) larger fat-tree topologies with 128 and 250 servers, (iv) a different workload with flow sizes uniformly distributed between 500KB to 5MB, representing background and storage traffic for RDMA, (v) the per-port buffer size varied between 60KB-480KB, (vi) varying other IRN parameters (increasing $RTO_{high}$ value by up to 4 times the default of 320$\mu$s, increasing the $N$ value for using $RTO_{low}$ to 10 and 15). We summarize our key observations 
here \crchange{and provide detailed results for each of these scenarios in Appendix \S A\notinarxiv{ of an extended report~\cite{irn-techrep}}.}

\paragraphb{Overall Results:} Across all of these experimental scenarios, we find that:

\paragraphi{(a)} IRN (without PFC) always performs better than RoCE (with PFC), with the performance improvement ranging from 6\% to 83\% across different cases. 

\paragraphi{(b)} When used without any congestion control, enabling PFC with IRN always degrades performance, with the maximum degradation across different scenarios being as high as 2.4$\times$.

\paragraphi{(c)} Even when used with Timely and DCQCN, enabling PFC with IRN often degrades performance (with the maximum degradation being 39\% for Timely and 20\% for DCQCN). Any improvement in performance due enabling PFC with IRN stays within 1.6\% for Timely and 5\% for DCQCN. 

\paragraphb{Some observed trends:} The drawbacks of enabling PFC with IRN:

\paragraphi{(a)} generally increase with increasing link utilization, as the negative impact of congestion spreading with PFC increases. 

\paragraphi{(b)} decrease with increasing bandwidths, as the relative cost of a round trip required to react to packet drops without PFC also increases.

\paragraphi{(c)} increase with decreasing buffer sizes due to more pauses and greater impact of congestion spreading. 

\crchange{We further observe that increasing $RTO_{high}$ or $N$ had a very small impact on our basic results, showing that IRN is not very sensitive to the specific parameter values.}

\subsubsection{Tail latency for small messages.}

We now look at the tail latency (or tail FCT) of the single-packet messages from our default scenario, which is another relevant metric in datacenters~\cite{timely}. Figure~\ref{fig:rpc-tail-latency} shows the tail CDF of this latency (from 90\%ile to 99.9\%ile), across different congestion control algorithms. Our key trends from \S\ref{sec:basicResults} hold even for this metric. 
This is because IRN (without PFC) is able to recover from single-packet message losses quickly due to the low $RTO_{low}$ timeout value. With PFC, these messages end up waiting in the queues for similar (or greater) duration due to pauses and congestion spreading. For all cases, IRN performs significantly better than RoCE.  

\begin{figure}[t]

\centering
 \renewcommand{\arraystretch}{0.1}
 \footnotesize

 \begin{subfigure}
 \centering
\includegraphics[width=0.45\textwidth]{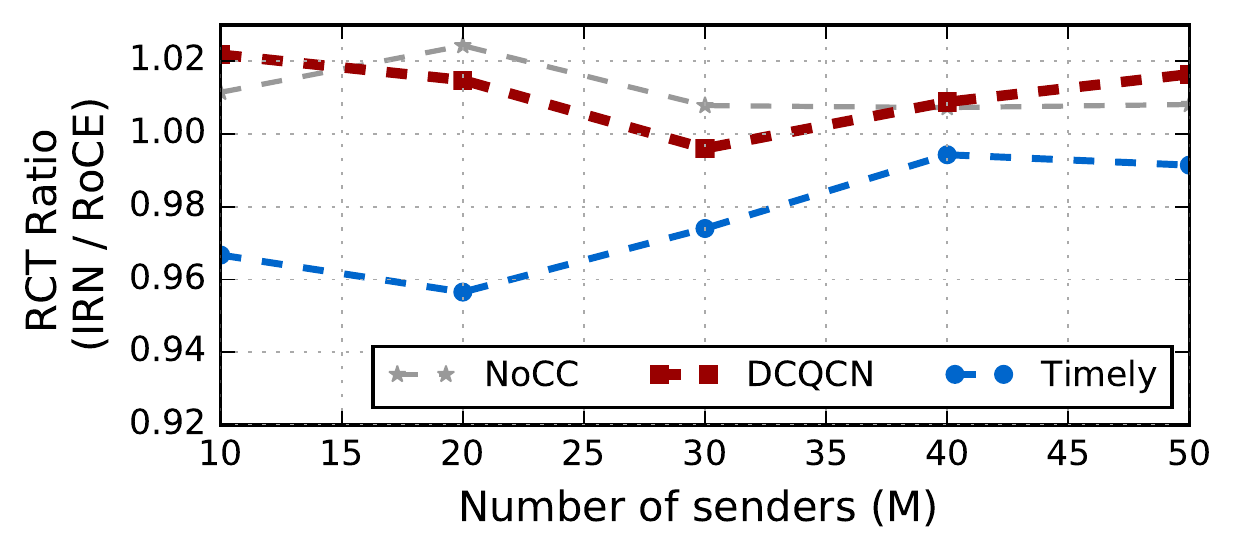}
 \end{subfigure}

 


\vspace{-10pt}

\caption{The figure shows the ratio of request completion time of incast with IRN (without PFC) over RoCE (with PFC) for varying degree of fan-ins across congestion control algorithms.}
\vspace{-5pt}

\label{fig:incast}
\end{figure}

\subsubsection{Incast.}
\label{sec:incast}

We now evaluate incast scenarios, both with and without cross-traffic. The incast workload without any cross traffic can be identified as the best case for PFC, since only valid congestion-causing flows are paused without unnecessary head-of-the-line blocking. 

\paragraphb{Incast without cross-traffic:} We simulate the incast workload on our default topology by striping 150MB of data across $M$ randomly chosen sender nodes that send it to a fixed destination node~\cite{pfabric}. We vary $M$ from 10 to 50. We consider the request completion time (RCT) as the metric for incast performance, which is when the last flow completes. For each $M$, we repeat the experiment 100 times and report the average RCT. Figure~\ref{fig:incast} shows the results, comparing IRN with RoCE. 
We find that the two have comparable performance: any increase in the RCT due to disabling PFC with IRN remained within 2.5\%. The results comparing IRN's performance with and without PFC looked very similar. We also varied our default incast setup by changing the bandwidths to 10Gbps and 100Gbps, and increasing the number of connections per machine. Any degradation in performance due to disabling PFC with IRN stayed within 9\%. 


\paragraphb{Incast with cross traffic:} 
In practice we expect incast to occur with other cross traffic in the network~\cite{timely, microsoftpfc}. We started an incast as described above with $M = 30$, along with our default case workload running at 50\% link utilization level. 
The incast RCT for IRN (without PFC) was always lower than RoCE (with PFC) by 4\%-30\% across the three congestion control schemes. 
For the background workload, the performance of IRN was better than RoCE by 32\%-87\% across the three congestion control schemes and the three metrics (\ie the average slowdown, the average FCT and the tail FCT). Enabling PFC with IRN generally degraded performance for both the incast and the cross-traffic by 1-75\% across the three schemes and metrics, and improved performance only for one case (incast workload with DCQCN by 1.13\%).

\subsubsection{Window-based congestion control.} 

We also implemented conventional window-based congestion control schemes such as TCP's AIMD and DCTCP~\cite{dctcp} with IRN and observed similar trends as discussed in \S\ref{sec:basicResults}. In fact, when IRN is used with TCP's AIMD, the benefits of disabling PFC were even stronger, because it exploits packet drops as a congestion signal, which is lost when PFC is enabled. 

\vspace{10pt}

\paragraphb{Summary:} Our key results \ie (1) IRN (without PFC) performs better than RoCE (with PFC), and (2) IRN does not require PFC, hold across varying realistic scenarios, congestion control schemes and performance metrics.

\begin{figure}[t]
\centering
 \renewcommand{\arraystretch}{0.1}
 \footnotesize

\renewcommand{\arraystretch}{1.0}
\newcolumntype{P}[1]{>{\centering\arraybackslash}p{#1}}
\setlength{\tabcolsep}{2pt}
\begin{tabular}{ccc}

\multicolumn{3}{c}{

 \begin{subfigure}
 \centering
\includegraphics[height=0.04\textwidth]{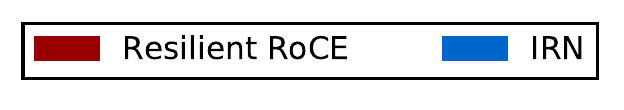}
 \end{subfigure} 

} \vspace{-10pt} \\

 \begin{subfigure}
 \centering
\includegraphics[width=0.125\textwidth,height=0.12\textwidth]{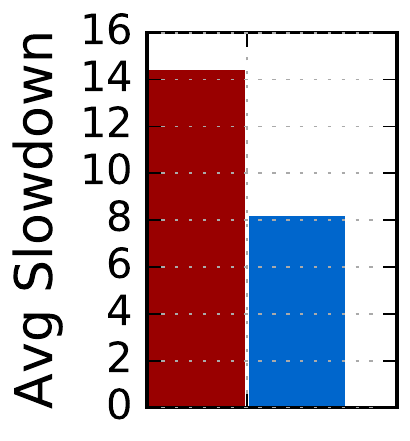}
 \end{subfigure} 
 
 &
 
  \begin{subfigure}
\centering
\includegraphics[width=0.125\textwidth,height=0.12\textwidth]{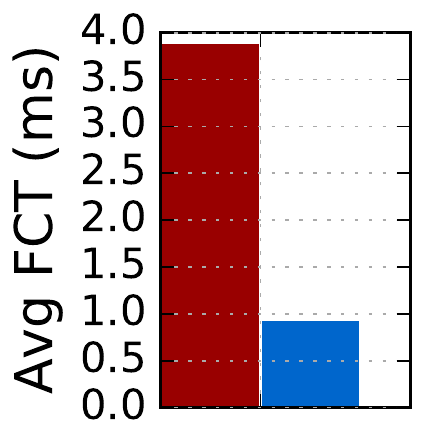}
\end{subfigure}

&

\begin{subfigure}
\centering
\includegraphics[width=0.125\textwidth,height=0.12\textwidth]{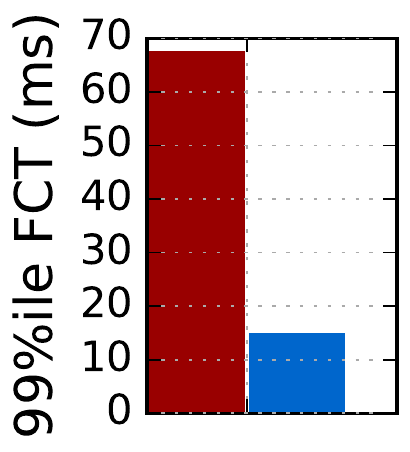}
\end{subfigure}


\end{tabular}

\caption{The figures compares resilient RoCE (RoCE+DCQCN without PFC) with IRN. 
}

\vspace{-10pt}
\label{fig:irn-vs-resRoCE}
\end{figure}

\crchange{
\subsection{Comparison with Resilient RoCE.} 
A recent proposal on Resilient RoCE~\cite{roceRocks} explores the use of DCQCN to avoid packet losses in specific scenarios, and thus eliminate the requirement for PFC. However, as observed previously in Figure \ref{fig:roce-withcc}, DCQCN may not always be successful in avoiding packet losses across all realistic scenarios with more dynamic traffic patterns and hence PFC (with its accompanying problems) remains necessary. Figure~\ref{fig:irn-vs-resRoCE} provides a direct comparison of IRN with Resilient RoCE. We find that IRN, even without any explicit congestion control, performs significantly better than Resilient RoCE, due to better loss recovery and BDP-FC. 
}

\subsection{Comparison with iWARP.}
\label{sec:iwarp-vs-irn}

\begin{figure}[t]
\centering
 \renewcommand{\arraystretch}{0.1}
 \footnotesize

\renewcommand{\arraystretch}{1.0}
\newcolumntype{P}[1]{>{\centering\arraybackslash}p{#1}}
\setlength{\tabcolsep}{2pt}
\begin{tabular}{ccc}

\multicolumn{3}{c}{
 \begin{subfigure}
 \centering
\includegraphics[height=0.04\textwidth]{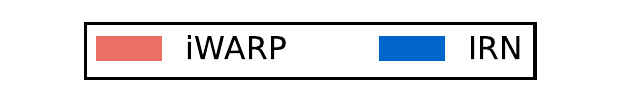}
 \end{subfigure} 

} \vspace{-10pt} \\

 \begin{subfigure}
 \centering
\includegraphics[width=0.125\textwidth,height=0.12\textwidth]{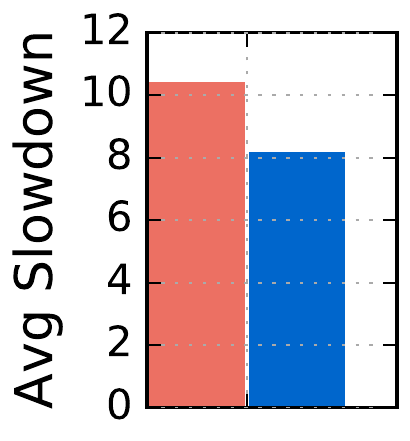}
 \end{subfigure} 
 
 &
 
  \begin{subfigure}
\centering
\includegraphics[width=0.125\textwidth,height=0.12\textwidth]{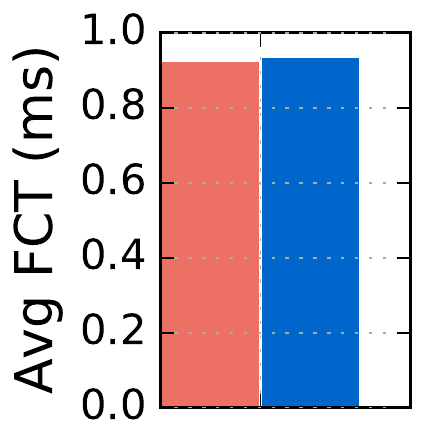}
\end{subfigure}

&

\begin{subfigure}
\centering
\includegraphics[width=0.125\textwidth,height=0.12\textwidth]{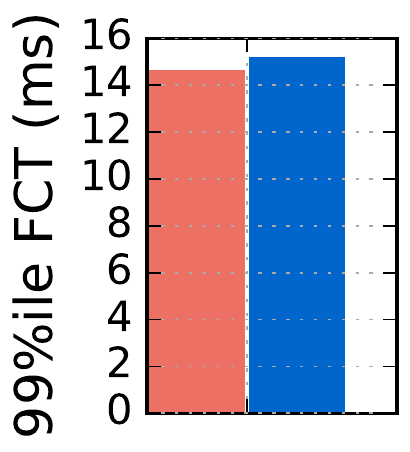}
\end{subfigure}


\end{tabular}

\caption{The figures compares iWARP's transport (TCP stack) with IRN. 
}

\vspace{-10pt}
\label{fig:irn-vs-iwarp}
\end{figure}

We finally explore whether IRN's simplicity over the TCP stack implemented in iWARP impacts performance. We compare IRN's performance (without any explicit congestion control) with full-blown TCP stack's, using INET simulator's in-built TCP implementation for the latter. Figure~\ref{fig:irn-vs-iwarp} shows the results for our default scenario. We find that absence of slow-start (with use of BDP-FC instead) results in 21\% smaller slowdowns with IRN and comparable average and tail FCTs. These results show that in spite of a simpler design, IRN's performance is better than full-blown TCP stack's, even without any explicit congestion control. Augmenting IRN with TCP's AIMD logic further improves its performance, resulting in 44\% smaller average slowdown and 11\% smaller average FCT as compared to iWARP. \crchange{ Furthermore, IRN's simple design allows it to achieve message rates comparable to current RoCE NICs with very little overheads (as evaluated in \S\ref{sec:feasibility-evaluation}). An iWARP NIC, on the other hand, can have up to 4$\times$ smaller message rate than a RoCE (\S\ref{sec:background}). Therefore, IRN provides a simpler and more performant solution than iWARP for eliminating RDMA's requirement for a lossless network.}

\section{Implementation Considerations}
\label{sec:irn-details}
We now discuss how one can incrementally update RoCE NICs to support IRN's transport logic, while maintaining the correctness of RDMA semantics as defined by the Infiniband RDMA specification~\cite{ibaspec}. Our implementation relies on extensions to RDMA's packet format, \eg introducing new fields and packet types.
These extensions are encapsulated within IP and UDP headers (as in RoCEv2) so they only effect the endhost behavior and not the network behavior (i.e. no changes are required at the switches). We begin with providing some relevant context about different RDMA operations before describing how IRN supports them.



\subsection{Relevant Context}

The two remote endpoints associated with an RDMA message transfer are called a \emph{requester} and a \emph{responder}. The interface between the user application and the RDMA NIC is provided by \emph{Work Queue Elements} or WQEs (pronounced as \emph{wookies}). The application posts a WQE for each RDMA message transfer, which contains the application-specified metadata for the transfer. It gets stored in the NIC while the message is being processed, and is expired upon message completion. The WQEs posted at the requester and at the responder NIC are called Request WQEs and Receive WQEs respectively. Expiration of a WQE upon message completion is followed by the creation of a \emph{Completion Queue Element} or a CQE (pronounced as \emph{cookie}), which signals the message completion to the user application. There are four types of message transfers supported by RDMA NICs:


\paragraphb{Write:} The requester \emph{writes} data to responder's memory. The data length, source and sink locations are specified in the Request WQE, and typically, no Receive WQE is required. However, Write-with-immediate operation requires the user application to post a Receive WQE that expires upon completion to generate a CQE (thus signaling Write completion at the responder as well).

\paragraphb{Read:} The requester \emph{reads} data from responder's memory. The data length, source and sink locations are specified in the Request WQE, and no Receive WQE is required.

\paragraphb{Send:} The requester \emph{sends} data to the responder. The data length and source location is specified in the Request WQE, while the sink location is specified in the Receive WQE.

\paragraphb{Atomic:} The requester reads and atomically updates the data at a location in the responder's memory, which is specified in the Request WQE. No Receive WQE is required. Atomic operations are restricted to single-packet messages.


\subsection{Supporting RDMA Reads and Atomics}
\label{sec:supportingReads}

IRN relies on per-packet ACKs for BDP-FC and loss recovery. RoCE NICs already support
per-packet ACKs for Writes and Sends. 
However, when doing Reads, the requester (which is the data sink) does not explicitly acknowledge the Read response packets.
IRN, therefore, introduces packets for \emph{read (N)ACKs} that are sent by a requester for each Read response packet. RoCE currently has eight unused opcode values available for the reliable connected QPs, and we use one of these for \emph{read (N)ACKs}. 
IRN also requires the Read responder (which is the data source) to implement timeouts. New timer-driven actions have been added to the NIC hardware implementation in the past~\cite{roceRocks}. Hence, this is not an issue. 

RDMA Atomic operations are treated similar to a single-packet RDMA Read messages. 

Our simulations from \S\ref{sec:losslessness-study} did not use ACKs for the RoCE (with PFC) baseline, modelling the extreme case of all Reads. Therefore, our results take into account the overhead of per-packet ACKs in IRN. 


\subsection{Supporting Out-of-order Packet Delivery}
\label{sec:oooDelivery}

One of the key challenges for implementing IRN is supporting out-of-order (OOO) packet delivery at the receiver -- current RoCE NICs simply discard OOO packets. A naive approach for handling OOO packet would be to store all of them in the NIC memory. The total number of OOO packets with IRN is bounded by the BDP cap (which is about 110 MTU-sized packets for our default scenario as described in \S\ref{sec:exptSettings})
\footnote{For QPs that only send single packet messages less than one MTU in size, the number of outstanding packets is limited to the maximum number of outstanding requests, 
which is typically smaller than the BDP cap~\cite{designGuidelines, herd}.}.
Therefore to support a thousand flows, a NIC would need to buffer 110MB of packets, which exceeds the memory capacity on most commodity RDMA NICs.

We therefore explore an alternate implementation strategy, where the NIC DMAs OOO packets directly to the final address in the application memory and keeps track of them using bitmaps (which are sized at BDP cap). This reduces NIC memory requirements from 1KB per OOO packet to only a couple of bits, but introduces some additional challenges that we address here. Note that partial support for OOO packet delivery was introduced in the Mellanox ConnectX-5 NICs to enable adaptive routing ~\cite{mlxcx5}. However, it is restricted to Write and Read operations. We improve and extend this design to support all RDMA operations with IRN.

We classify the issues due to out-of-order packet delivery into four categories.

\subsubsection{First packet issues.}

For some RDMA operations, critical information is carried in the first packet of a message, which is required to process other packets in the message. Enabling OOO delivery, therefore, requires that some of the information in the first packet be carried by all packets. 

In particular, the RETH header (containing the remote memory location) is carried only by the first packet of a Write message. IRN requires adding it to every packet.

\subsubsection{WQE matching issues.} 
\label{sec:wqematching}

Some operations require every packet that arrives to be matched with its corresponding WQE at the responder. This is done implicitly for in-order packet arrivals. However, this implicit matching breaks with OOO packet arrivals. A work-around for this is assigning explicit WQE sequence numbers, that get carried in the packet headers and can be used to identify the corresponding WQE for each packet. IRN uses this workaround for the following RDMA operations:

\paragraphi{Send and Write-with-immediate:} It is required that Receive WQEs be consumed by Send and Write-with-immediate requests in the same order in which they are posted. Therefore, with IRN every Receive WQE, and every Request WQE for these operations, maintains a \emph{recv\_WQE\_SN} that indicates the order in which they are posted. This value is carried in all Send packets and in the last Write-with-Immediate packet,~\footnote{A Receive WQE is consumed only by the last packet of a Write-with-immediate message, and is required to process all packets for a Send message.} and is used to identify the appropriate Receive WQE. IRN also requires the Send packets to carry the relative offset in the packet sequence number, which is used to identify the precise address when placing data. 

\paragraphi{Read/Atomic:} The responder cannot begin processing a Read/Atomic request $R$, until all packets expected to arrive before $R$ have been received. Therefore, an OOO Read/Atomic Request packet needs to be placed in a Read WQE buffer at the responder (which is already maintained by current RoCE NICs). With IRN, every Read/Atomic Request WQE maintains a \emph{read\_WQE\_SN}, that is carried by all Read/Atomic request packets and allows identification of the correct index in this Read WQE buffer. 

\subsubsection{Last packet issues.} 
For many RDMA operations, critical information is carried in last packet, which is required to complete message processing. Enabling OOO delivery, therefore, requires keeping track of such last packet arrivals and storing this information at the endpoint (either on NIC or main memory), until all other packets of that message have arrived. We explain this in more details below.

A RoCE responder maintains a \emph{message sequence number (MSN)} which gets incremented when the last packet of a Write/Send message is received or when a Read/Atomic request is received. This MSN value is sent back to the requester in the ACK packets and is used to expire the corresponding Request WQEs. 
The responder also expires its Receive WQE when the last packet of a Send or a Write-With-Immediate message is received and generates a CQE. The CQE is populated with certain meta-data about the transfer, which is carried by the last packet. 
IRN, therefore, needs to ensure that the completion signalling mechanism works correctly even when the last packet of a message arrives before others. For this, an IRN responder maintains a 2-bitmap, which in addition to tracking whether or not a packet $p$ has arrived, also tracks whether it is the last packet of a message that will trigger (1) an MSN update and (2) in certain cases, a Receive WQE expiration that is followed by a CQE generation. These actions are triggered only after all packets up to $p$ have been received. For the second case, the \emph{recv\_WQE\_SN} carried by $p$ (as discussed in \S\ref{sec:wqematching}) can identify the Receive WQE with which the meta-data in $p$ needs to be associated, thus enabling a \emph{premature CQE} creation. The premature CQE can be stored in the main memory, until it gets delivered to the application after all packets up to $p$ have arrived. 

\subsubsection{Application-level Issues.}




Certain applications (for example FaRM~\cite{farm}) rely on polling the last packet of a Write message to detect completion, which is incompatible with OOO data placement. 
This polling based approach violates the RDMA specification (Sec o9-20~\cite{ibaspec}) and is more expensive than officially supported methods (FaRM~\cite{farm} mentions moving on to using the officially supported Write-with-Immediate method in the future for better scalability).
IRN's design provides all of the Write completion guarantees as per the RDMA specification. \crchange{This is discussed in more details in Appendix ~\S B\notinarxiv{ of the extended report~\cite{irn-techrep}}.} 

OOO data placement can also result in a situation where data written to a particular memory location is overwritten by a restransmitted packet from an older message.
Typically, applications using distributed memory frameworks assume relaxed memory ordering and use application layer \emph{fences} whenever strong memory consistency is required~\cite{appfence1, appfence2}. 
Therefore, both iWARP and Mellanox ConnectX-5, in supporting OOO data placement, expect the application to deal with the potential memory over-writing issue and do not handle it in the NIC or the driver. IRN can adopt the same strategy. Another alternative is to deal with this issue in the driver, by enabling the fence indicator for a newly posted request that could potentially overwrite an older one. 

\subsection{Other Considerations}
\label{sec:otherdetails}


Currently, the packets that are sent and received by a requester use the same packet sequence number ($PSN$) space. This interferes with loss tracking and BDP-FC. IRN, therefore, splits the $PSN$ space into two different ones (1) $sPSN$ to track the request packets sent by the requester, and (2) $rPSN$ to track the response packets received by the requester. This decoupling remains transparent to the application and is compatible with the current RoCE packet format. IRN can also support shared receive queues and send with invalidate operations and is compatible with use of end-to-end credit. \crchange{We provide more details about these in Appendix \S B\notinarxiv{ of the extended report~\cite{irn-techrep}}.}
\section{Evaluating Implementation Overheads}
\label{sec:feasibility-evaluation}

We now evaluate IRN's implementation overheads over current RoCE NICs along the following three dimensions: in \S\ref{sec:stateOverhead}, we do a comparative analysis of IRN's memory requirements; in \S\ref{sec:fpgaSynthesis}, we evaluate the overhead for implementing IRN's packet processing logic by synthesizing it on an FPGA;  and in \S\ref{sec:perfOverhead}, we evaluate, via simulations, how IRN's implementation choices impact end-to-end performance. 

\subsection{NIC State overhead due to IRN}
\label{sec:stateOverhead}

\noindent Mellanox RoCE NICs support several MBs of cache to store various metadata including per-QP and per-WQE contexts. 
The additional state that IRN introduces consumes a total of
only 3-10\% of the current NIC cache for a couple of thousands of QPs and tens of thousands of WQEs, even when considering large 100Gbps links. We present a breakdown this additional state below.


\paragraphb{Additional Per-QP Context:}

\paragraphi{State variables:} IRN needs 52 bits of additional state for its transport logic: 24 bits each to track the packet sequence to be retransmitted and the recovery sequence, and 4 bits for various flags.
Other per-flow state variables needed for IRN's transport logic (e.g., expected sequence number) 
are already maintained by current RoCE NICs. Hence, the per-QP overhead is 104 bits (52 bits each at the requester and the responder). Maintaining a timer at the responder for Read timeouts and a variable to track in-progress Read requests in the Read WQE buffer adds another 56 bits to the responder leading to a total of 160 bits of additional per-QP state with IRN. 
For context, RoCE NICs currently maintain a few thousands of bits 
per QP for various state variables.

\paragraphi{Bitmaps:} IRN requires five BDP-sized bitmaps: two at the responder for the 2-bitmap to track received packets, one at the requester to track the Read responses, one each at the requester and responder for tracking selective acks. Assuming each bitmap to be 128 bits (\ie sized to fit the BDP cap for a network with bandwidth 40Gbps and a two-way propagation delay of up to 24$\mu$s, typical in today's datacenter topologies~\cite{timely}),
IRN would require a total of 640 bits per QP for bitmaps.
This is much less than the total size of bitmaps maintained by a QP for the OOO support in Mellanox ConnectX-5 NICs. 

\paragraphi{Others:} Other per-QP meta-data that is needed by an IRN driver when a WQE is posted (e.g  counters for assigning WQE sequence numbers) or expired (e.g. premature CQEs) can be stored directly in the main memory and do not add to the NIC memory overhead.

\paragraphb{Additional Per-WQE Context:} As described in \S\ref{sec:irn-details}, IRN maintains sequence numbers for certain types of WQEs. This adds 3 bytes to the per-WQE context which is currently sized at 64 bytes. 

\paragraphb{Additional Shared State:} IRN also maintains some additional variables (or parameters) that are shared across QPs. This includes the BDP cap value, the $RTO_{low}$ value, and $N$ for $RTO_{low}$, which adds up to a total of only 10 bytes. 







\subsection{IRN's packet processing overhead}
\label{sec:fpgaSynthesis}

We evaluate the implementation overhead due to 
IRN's per-packet processing logic, which requires various bitmap manipulations.
The logic for other changes that IRN makes -- e.g., adding header extensions to packets, premature CQE generation, etc. -- are already implemented in RoCE NICs and can be easily extended for IRN.

We use Xilinx Vivado Design Suite 2017.2 ~\cite{vivado} to do a high-level synthesis of the four key packet processing modules (as described below), targeting the Kintex Ultrascale XCKU060 FPGA which is supported as a bump-on-the-wire on the Mellanox Innova Flex 4 10/40Gbps NICs~\cite{mlxInnovaFlex}.

\subsubsection{Synthesis Process.}
\label{sec:synthmodules}

To focus on the \emph{additional} packet processing complexity due to IRN, our implementation for the four modules is \emph{stripped-down}. More specifically, each module receives the relevant packet metadata and the QP context 
as streamed inputs, relying on a RoCE NIC's existing implementation to parse the packet headers and retrieve the QP context from the NIC cache (or the system memory, in case of a cache miss). The updated QP context is passed as streamed output from the module, along with other relevant module-specific outputs as described below. 

\paragraphi{(1) receiveData:} Triggered on a packet arrival, it outputs the relevant information required to generate an ACK/NACK packet and the number of Receive WQEs to be expired, along with the updated QP context (e.g. bitmaps, expected sequence number, MSN).  

\paragraphi{(2) txFree:} Triggered when the link's Tx is free for the QP to transmit, it outputs the sequence number of the packet to be (re-)transmitted and the updated QP context (e.g. next sequence to transmit). During loss-recovery, it also performs a look ahead by searching the SACK bitmap for the next packet sequence to be retransmitted.

\paragraphi{(3) receiveAck:} Triggered when an ACK/NACK packet arrives, 
it outputs the updated QP context (e.g. SACK bitmap, last acknowledged sequence).

\paragraphi{(4) timeout:} If triggered when the timer expires using $RTO_{low}$ value (indicated by a flag in the QP context), it checks if the condition for using $RTO_{low}$ holds. If not, it does not take any action and sets an output flag to extend the timeout to $RTO_{high}$. In other cases, it executes the timeout action and returns the updated QP context. Our implementation relies on existing RoCE NIC's support for setting timers, with the $RTO_{low}$ value being used by default, unless explicitly extended. 




The bitmap manipulations in the first three modules account for most of the complexity in our synthesis. Each bitmap was implemented as a ring buffer, using an arbitrary precision variable of 128 bits, with the \emph{head} corresponding to the expected sequence number at the receiver (or the cumulative acknowledgement number at the sender). The key bitmap manipulations required by IRN can be reduced to the following three categories of known operations: (i) \emph{finding first zero}, to find the next expected sequence number in \emph{receiveData} 
and the next packet to retransmit in \emph{txFree} (ii) \emph{popcount} to compute the increment in MSN and the number of Receive WQEs to be expired in \emph{receiveData}, (iii) \emph{bit shifts} to advance the bitmap \emph{heads} in \emph{receiveData} and \emph{receiveAck}. We optimized the first two operations by dividing the bitmap variables into chunks of 32 bits and operating on these chunks in parallel. 

We validated the correctness of our implementation by generating input event traces for each synthesized module from the  simulations described in \S\ref{sec:losslessness-study} and passing them as input in the test bench used for RTL verification by the Vivado Design Suite. The output traces, thus, generated were then matched with the corresponding output traces obtained from the simulator. We also used the Vivado HLS tool to export our RTL design to create IP blocks for our modules.



\begin{table}[t]
\centering
\scriptsize
\renewcommand{\arraystretch}{1.25}
\newcolumntype{P}[1]{>{\centering\arraybackslash}p{#1}}
\setlength{\tabcolsep}{3.2pt}
\begin{tabular}[b]{|P{1.5cm}|P{0.8cm}|P{0.8cm}|P{1cm}|P{1.5cm}|}
\hline
\multirow{2}{*}{\parbox{1.5cm}{\centering \textbf{Module Name}}} & 
\multicolumn{2}{c|}{\textbf{Resource Usage}} & 
\textbf{Max} &
\textbf{Min} \\
\cline{2-3}
& FF & LUT & \textbf{Latency} & \textbf{Throughput} \\
\hline
\emph{receiveData} & 0.62\% & 1.93\% & 16.5 ns & 45.45 Mpps \\
\emph{txFree} & 0.32\% & 0.95\% & 15.9 ns & 47.17  Mpps \\
\emph{receiveAck} & 0.4\% & 1.05\% & 15.96 ns & 46.99  Mpps \\
\emph{timeout} & 0.01\% & 0.08\% & $<$6.3 ns & 318.47  Mpps \\ 
\hline
\multicolumn{5}{|l|}{\textbf{Total Resource Usage: 1.35\% FF and 4.01\% LUTs}} \\
\hline
\multicolumn{5}{|l|}{\textbf{Min Bottleneck Tpt: 45.45Mpps}} \\
\hline

\end{tabular}
\caption{Performance and resource usage for different packet processing modules on Xilinx Kintex Ultrascale KU060 FPGA. }
\vspace{-10pt}
\label{tab:fpgaSynthesisReport}
\end{table}

\subsubsection{Synthesis Results.} Our FPGA synthesis report has been summarized in Table~\ref{tab:fpgaSynthesisReport} and discussed below.

\paragraphb{Resource Usage:} The second and third columns in Table~\ref{tab:fpgaSynthesisReport} report the percentage of flip-flops (FF) and look-up tables (LUT) used for the four modules (no BRAM or DSP48E units were consumed). We find that each of IRN's packet processing modules consume less than 1\% FFs and 2\% LUTs (with a total of 1.35\% FFs and 4\% LUTs consumed). Increasing the bitmap size to support 100Gbps links consumed a total of 2.66\% of FFs and 9.5\% of LUTs on the same device (though we expect the relative resource usage to be smaller on a higher-scale device designed for 100Gbps links).  

\paragraphb{Performance:} The third and fourth column in Table~\ref{tab:fpgaSynthesisReport} report the worst-case latency and throughput respectively for each module. \footnote{The worst-case throughput was computed by dividing the clock frequency with the maximum initiation interval, as reported by the Vivado HLS synthesis tool~\cite{xilinx-ug-902}.} 
The latency added by each module is at most only 16.5ns. The \emph{receiveData} module (requiring more complex bitmap operations) had the lowest throughput of 45.45Mpps. This is high enough to sustain a rate of 372Gbps for MTU-sized packets. It is also higher than the maximum rate of 39.5Mpps that we observed on Mellanox MCX416A-BCAT RoCE NIC across different message sizes (2 bytes - 1KB), after applying various optimizations such as batching and using multiple queue-pairs. A similar message rate was observed in prior work~\cite{herd}. Note that we did not use pipelining within our modules, which can further improve throughput. 



While we expect IRN to be implemented on the ASIC integrated with the existing RoCE implementation, we believe that the modest resources used on an FPGA board supported as an \emph{add-on} in recent RDMA-enabled NICs, provides some intuition about the feasibility of the changes required by IRN. Also, note that the results reported here are far from the optimal results that can be achieved on an ASIC implementation due to two sources of sub-optimality: (i) using HLS for FPGA synthesis has been found to be up to 2$\times$ less optimal than directly using Verilog~\cite{clicknp} and (ii) FPGAs, in general, are known to be less optimal than ASICs.


\begin{figure*}[t]

\centering
 \renewcommand{\arraystretch}{0.1}
 \footnotesize

\begin{tabular}{ccc}

\multicolumn{3}{c}{

 \begin{subfigure}
 \centering
\includegraphics[height=0.04\textwidth]{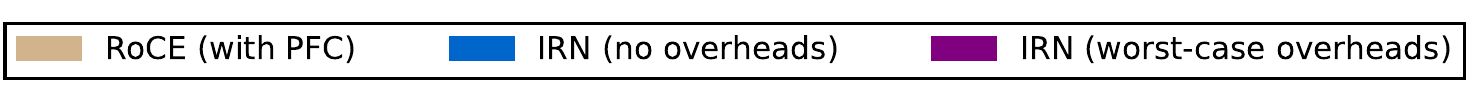}
 \end{subfigure} 

} \vspace{-10pt} \\

 \begin{subfigure}
\centering
\includegraphics[width=0.3\textwidth]{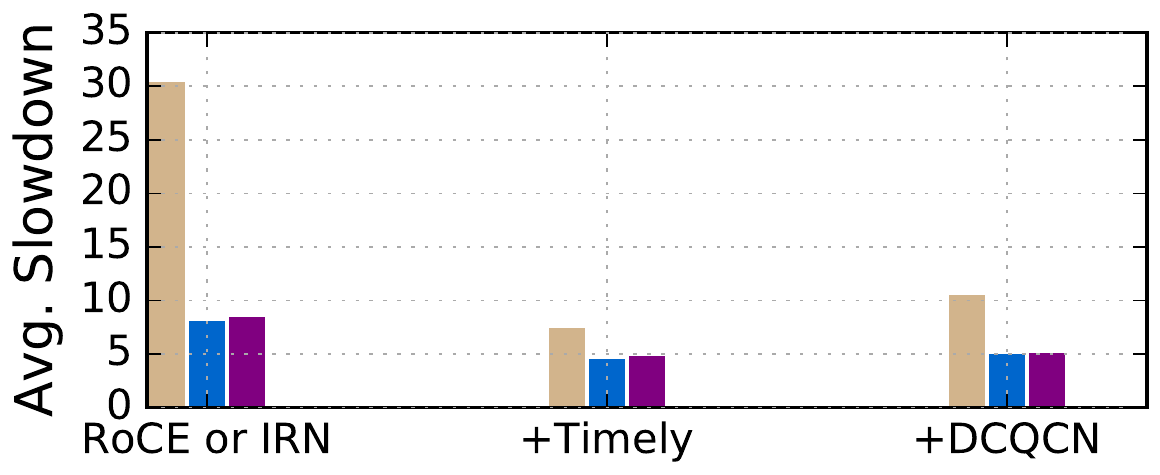}
 \end{subfigure}
 
 &
 
\begin{subfigure}
\centering
\includegraphics[width=0.3\textwidth]{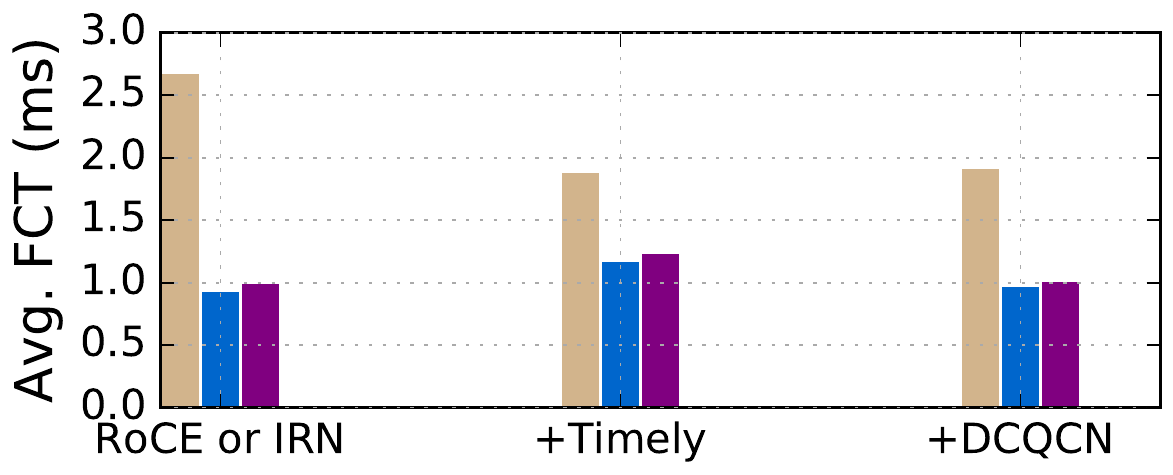}
\end{subfigure}

&

\begin{subfigure}
\centering
\includegraphics[width=0.3\textwidth]{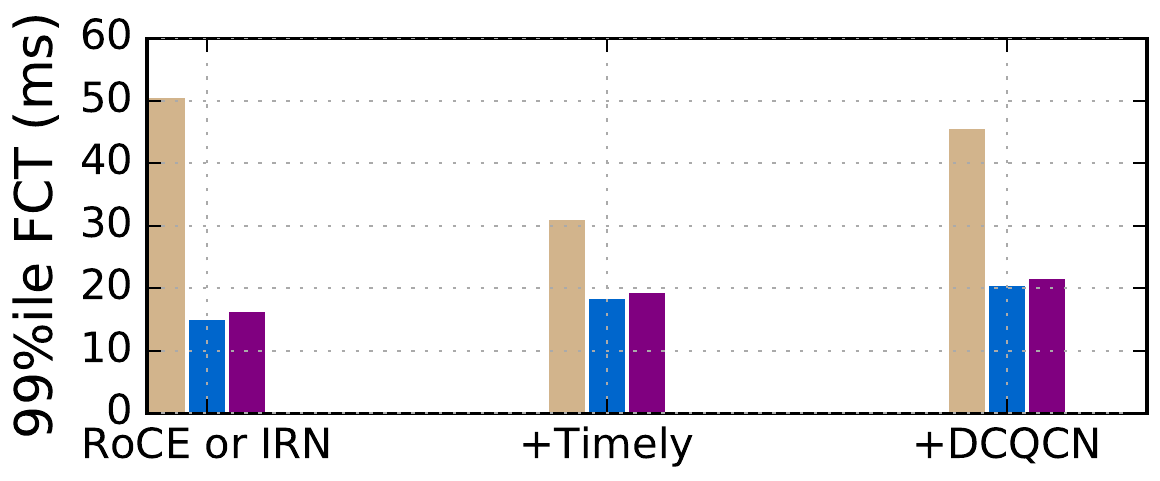}
\end{subfigure}

\end{tabular}

\caption{The figures show the performance of IRN with worse case overheads, comparing it with IRN without any overheads and with RoCE for our default case scenario.}
\vspace{-10pt}
\label{fig:irn-with-overheads}
\end{figure*}

\subsection{Impact on end-to-end performance}
\label{sec:perfOverhead}

We now evaluate how IRN's implementation overheads impact the end-to-end performance. We identify the following two implementation aspects that could potentially impact end-to-end performance and model these in our simulations.

\paragraphb{Delay in Fetching Retransmissions:} While the regular packets sent by a RoCE NIC are typically pre-fetched, we assume that the DMA request for retransmissions is sent only after the packet is identified as lost (i.e. when loss recovery is triggered or when a look-ahead is performed). 
The time taken to fetch a packet over PCIe is typically between a few hundred nanoseconds to $<$2$\mu$s~\cite{xilinxPCIeLatency, senic}. 
We set a worst-case retransmission delay of 2$\mu$s for every retransmitted packet i.e. the sender QP is allowed to retransmit a packet only after 2$\mu$s have elapsed since the packet was detected as lost. 

\paragraphb{Additional Headers:} As discussed in \S\ref{sec:irn-details}, some additional headers are needed in order to DMA the packets directly to the application memory, of which, the most extreme case is the 16 bytes of RETH header added to every Write packet. Send data packets have an extra header of 6 bytes, while Read responses do not require additional headers. We simulate the worst-case scenario of all Writes with every packet carrying 16 bytes additional header. 

\paragraphb{Results:} Figure~\ref{fig:irn-with-overheads} shows the results for our default scenario after modeling these two sources of worst-case overheads. We find that they make little difference to the end-to-end performance (degrading the performance by 4-7\% when compared to IRN without overheads). The performance remains 35\%-63\% better than our baseline of RoCE (with PFC). We also verified that the retransmission delay of 2$\mu$s had a much smaller impact on end-to-end performance (2$\mu$s is very small compared to the network round-trip time taken to detect a packet loss and to recover from it, which could be of the order of a few hundred microseconds). The slight degradation in performance observed here can almost entirely by attributed to the additional 16 bytes header in every packet. Therefore, we would expect the performance impact to be even smaller when there is a mix of Write and Read workloads. 


\subsection{Summary}

Our analysis shows that IRN is well within the limits of feasibility, with small chip area and NIC memory requirements and minor bandwidth overhead. We also validated our analysis through extensive discussions with two commercial NIC vendors (including Mellanox); both vendors confirmed that the IRN design can be easily implemented on their hardware NICs. Inspired by the   results presented in this paper, Mellanox is  implementing a version of IRN in their next release. 


\section{Discussion and Related Work}
\label{sec:discussion}

\paragraphb{Backwards Compatibility:} We briefly sketch one possible path to incrementally deploying IRN. We envision that NIC vendors will manufacture NICs that support dual RoCE/IRN modes. The use of IRN can be negotiated between two endpoints via the RDMA connection manager, with the NIC falling back to RoCE mode if the remote endpoint does not support IRN. (This is similar to what was used in moving from RoCEv1 to RoCEv2.) Since IRN performs well both with or without PFC, network operators can continue to run PFC until all their endpoints have been upgraded to support IRN at which point PFC can be permanently disabled. During the interim period, hosts can 
communicate using either RoCE or IRN, with no loss in performance.

\paragraphb{Reordering due to load-balancing:} Datacenters today use ECMP for load balancing~\cite{microsoftpfc}, that maintains ordering within a flow. IRN's OOO packet delivery support also allows for other load balancing schemes that may cause packet reordering within a flow~\cite{drill, packetSpraying}. IRN's loss recovery mechanism can be made more robust to reordering by triggering loss recovery only after a certain threshold of NACKs are received. 

\paragraphb{Other hardware-based loss recovery:} MELO~\cite{melo}, a recent scheme developed in parallel to IRN, proposes an alternative design for hardware-based selective retransmission, where out-of-order packets are buffered in an off-chip memory. Unlike IRN, MELO only  targets PFC-enabled environments with the aim of greater robustness to random losses caused by failures. As such, MELO is orthogonal to our main focus which is showing that PFC is unnecessary. Nonetheless, the existence of alternate designs such as MELO's further corroborates the feasibility of implementing better loss recovery on NICs.


 \paragraphb{HPC workloads:} The HPC community has long been a strong supporter of losslessness. This is primarily because HPC clusters are smaller with more controlled traffic patterns, and hence the negative effects of providing losslessness (such as congestion spreading and deadlocks) are rarer. PFC's issues are exacerbated on larger scale clusters~\cite{microsoftpfc, dcqcn, timely, pfcdeadlocks1, pfcdeadlocks2}. 
 
\paragraphb{Credit-based Flow Control:} Since the focus of our work was RDMA deployment over Ethernet, our experiments used PFC.
Another approach to losslessness, used by Infiniband, is credit-based flow control, where the downlink sends credits to the uplink when it has sufficient buffer capacity. Credit-based flow control suffers from the same performance issues as PFC: head-of-the-line blocking, congestion spreading, the potential for deadlocks, {\it etc}. We, therefore, believe that our observations from \S\ref{sec:losslessness-study} can be applied to credit-based flow control as well.

\section{Acknowledgement}

We would like to thank Amin Tootoonchian, Anirudh Sivaraman, and Emmanuel Amaro for the helpful discussions and feedback on some of the implementation specific aspects of this work, and Brian Hausauer for his detailed feedback on an earlier version of this paper. We are also thankful to Nandita Dukkipati and Amin Vahdat for the useful discussions in the early stages of this work. We would finally like to thank our anonymous reviewers for their feedback which helped us in improving the paper, and our shepherd Srinivasan Seshan who helped shape the final version of this paper. This work was supported in parts by a Google PhD Fellowship and by Mellanox, Intel and the National Science Foundation under Grant No. 1704941 and 1619377. 

\normalsize{}

\section*{Appendix}
 
\begin{appendices}
\section{Evaluating IRN Across Varying Experimental Scenarios}
\label{app:detailed-robustness}


Here we present detailed results for our experiments across different scenarios (summarized in \S 4.4.1 of the main paper). The results are presented in a tabular format, and include for each the three metrics we consider (\ie average slowdown, average FCT and tail FCT):
(i) the absolute value of the metric with IRN, (ii) the ratio of the metric for IRN over IRN + PFC (if this ratio is smaller than or close enough to 1, \ie less than 1.1, then it shows that IRN does not require PFC), and (iii) the ratio of the metric for IRN over RoCE + PFC (if this ratio is less than 1, then it shows that IRN without PFC performs better than RoCE with PFC). We present these set of results for all three cases that we consider in the main paper: (a) without any explicit congestion control, (b) with Timely, and (c) with DCQCN.


\begin{table*}[t]
\centering
\scriptsize
\renewcommand{\arraystretch}{1.5}
\newcolumntype{P}[1]{>{\centering\arraybackslash}p{#1}}
\begin{tabular}[b]{|P{1.5cm}|P{1.5cm}|P{0.7cm}P{0.7cm}P{0.7cm}|P{0.7cm}P{0.7cm}P{0.7cm}|P{0.7cm}P{0.7cm}P{0.7cm}|}
\hline
\multirow{2}{*}{\textbf{Link Util.}}
& & \multicolumn{3}{c|}{\textbf{IRN}} & \multicolumn{3}{c|}{\textbf{IRN + Timely}} & \multicolumn{3}{c|}{\textbf{IRN + DCQCN}} \\
\cline{3-11} 
& & Avg Slowdown & Avg FCT & 99\%ile FCT & Avg Slowdown & Avg FCT & 99\%ile FCT & Avg Slowdown & Avg FCT & 99\%ile FCT \\
\hline
\multirow{3}{*}{30\%} 
& IRN
& 1.85 & 0.0002 & 0.0024
& 1.70 & 0.0003 & 0.0059
& 1.69 & 0.0002 & 0.0029
\\
& $\frac{\text{IRN}}{\text{IRN+PFC}}$ 
&  0.990 & 1.000 & 0.995
&  0.996 & 0.993 & 0.999
&  1.003 & 1.005 & 1.010 
\\
& $\frac{\text{IRN}}{\text{RoCE+PFC}}$ 
& 0.335 & 0.743 & 0.852
& 0.890 & 0.756 & 0.789
&  0.904 & 0.909 & 0.936
\\
\hline
\multirow{3}{*}{50\%} 
& IRN
& 3.38 & 0.0003 & 0.0046 
& 2.55 & 0.0006 & 0.0101
& 2.65 & 0.0004 & 0.0057
\\
& $\frac{\text{IRN}}{\text{IRN+PFC}}$ 
&  0.868 & 0.931 & 0.930
&  0.996 & 0.993 & 0.989  
&  1.001 & 1.000 & 1.015
\\
& $\frac{\text{IRN}}{\text{RoCE+PFC}}$ 
& 0.186 & 0.344 & 0.335
& 0.801 & 0.755 & 0.776  
& 0.746 & 0.771 & 0.626
\\
\hline
\multirow{3}{*}{\emph{70\%}} 
& IRN
&  8.24 & 0.0009 & 0.0153
&  4.73 & 0.0012 & 0.0185
&  5.19 & 0.0010 & 0.0207
\\
& $\frac{\text{IRN}}{\text{IRN+PFC}}$ 
& 0.513 & 0.640 & 0.612
& 0.995 & 0.976 & 0.968
& 1.009 & 1.005 & 0.966
\\
& $\frac{\text{IRN}}{\text{RoCE+PFC}}$ 
& 0.269 & 0.350 & 0.301
& 0.626 & 0.625 & 0.594
& 0.484 & 0.509 & 0.453
\\
\hline
\multirow{3}{*}{90\%} 
& IRN
& 14.03 & 0.0019 & 0.0334
&  7.84 & 0.0019 & 0.0321
&  8.13 & 0.0019 & 0.0464
\\
& $\frac{\text{IRN}}{\text{IRN+PFC}}$ 
& 0.483 & 0.572 & 0.484
& 0.930 & 0.868 & 0.761
& 1.004 & 0.990 & 0.964
\\
& $\frac{\text{IRN}}{\text{RoCE+PFC}}$ 
& 0.359 & 0.475 & 0.468
& 0.501 & 0.518 & 0.487
& 0.534 & 0.592 & 0.697
\\
\hline

\end{tabular}
\caption{Robustness of IRN with varying average link utilization levels}
\vspace{-5pt}
\label{tab:util-details}
\end{table*}

\begin{table*}[t]
\centering
\scriptsize
\renewcommand{\arraystretch}{1.5}
\newcolumntype{P}[1]{>{\centering\arraybackslash}p{#1}}
\begin{tabular}[b]{|P{1.5cm}|P{1.5cm}|P{0.7cm}P{0.7cm}P{0.7cm}|P{0.7cm}P{0.7cm}P{0.7cm}|P{0.7cm}P{0.7cm}P{0.7cm}|}
\hline
\multirow{2}{*}{\textbf{Bandwidth}}
& & \multicolumn{3}{c|}{\textbf{IRN}} & \multicolumn{3}{c|}{\textbf{IRN + Timely}} & \multicolumn{3}{c|}{\textbf{IRN + DCQCN}} \\
\cline{3-11} 
& & Avg Slowdown & Avg FCT & 99\%ile FCT & Avg Slowdown & Avg FCT & 99\%ile FCT & Avg Slowdown & Avg FCT & 99\%ile FCT \\
\hline
\multirow{3}{*}{10Gbps} 
& IRN
& 9.41 & 0.0035 & 0.0595 
& 5.45 & 0.0039 & 0.0616
& 5.88 & 0.0040 & 0.0809 
\\
& $\frac{\text{IRN}}{\text{IRN+PFC}}$ 
&  0.371 & 0.523 & 0.458 
&  0.967 & 0.924 & 0.903  
& 0.973 & 0.929 & 0.954 
\\
& $\frac{\text{IRN}}{\text{RoCE+PFC}}$ 
& 0.170 & 0.309 & 0.274
& 0.720 & 0.637 & 0.620
& 0.695 & 0.681 & 0.688
\\
\hline
\multirow{3}{*}{\emph{40Gbps}} 
& IRN
&  8.24 & 0.0009 & 0.0153
&  4.73 & 0.0012 & 0.0185
&  5.19 & 0.0010 & 0.0207
\\
& $\frac{\text{IRN}}{\text{IRN+PFC}}$ 
& 0.513 & 0.640 & 0.612
& 0.995 & 0.976 & 0.968
& 1.009 & 1.005 & 0.966
\\
& $\frac{\text{IRN}}{\text{RoCE+PFC}}$ 
& 0.269 & 0.350 & 0.301
& 0.626 & 0.625 & 0.594
& 0.484 & 0.509 & 0.453
\\
\hline
\multirow{3}{*}{100Gbps} 
& IRN
&  7.92 & 0.0004 & 0.0064
& 4.84 & 0.0006 & 0.0091
& 5.44 & 0.0007 & 0.0187
\\
& $\frac{\text{IRN}}{\text{IRN+PFC}}$ 
& 0.629 & 0.728 & 0.705
& 0.988 & 0.986 & 0.964
& 1.011 & 1.022 & 1.051
\\
& $\frac{\text{IRN}}{\text{RoCE+PFC}}$ 
&  0.408 & 0.476 & 0.413
& 0.385 & 0.489 & 0.432
& 0.424 & 0.597 & 0.658
\\
\hline

\end{tabular}
\caption{Robustness of IRN with varying average link bandwidth}
\vspace{-5pt}
\label{tab:bandwidth-details}
\end{table*}

\subsection{Varying link utilization levels.} Table~\ref{tab:util-details} shows the robustness of our basic results as the link utilization level is varied from 30\% to 90\%. We find that as the link utilization increases, both the ratios for each metric (\ie IRN over IRN + PFC and IRN over RoCE + PFC) decrease, indicating that IRN (without PFC) performs increasingly better than both IRN + PFC and RoCE + PFC. This follows from the fact that the drawbacks of using PFC increases at higher link utilization due to increased congestion spreading.

\subsection{Varying bandwidth} Table~\ref{tab:bandwidth-details} shows how our results vary as the bandwidth is changed from our default of 40Gbps to a lower value of 10Gbps and a higher value of 100Gbps. Here we find that as the bandwidth increases, the relative cost of a round trip required to react to packet drops without PFC also increases, thus reducing the performance gap between IRN (without PFC) and the two PFC enabled cases.

\subsection{Varying the scale of topology.} Table~\ref{tab:topoSize-details} shows the robustness of our basic results as the scale of the topology is increased from our default of 6 port switches with 54 servers to 8 and 10 port switches with 128 and 250 servers respectively. Our trends remain roughly similar as we scale up the topology beyond our default set up. 

\begin{table*}[t]
\centering
\scriptsize
\renewcommand{\arraystretch}{1.5}
\newcolumntype{P}[1]{>{\centering\arraybackslash}p{#1}}
\begin{tabular}[b]{|P{1.5cm}|P{1.5cm}|P{0.7cm}P{0.7cm}P{0.7cm}|P{0.7cm}P{0.7cm}P{0.7cm}|P{0.7cm}P{0.7cm}P{0.7cm}|}
\hline
\multirow{2}{*}{\parbox{1.5cm}{\centering \textbf{Scale-out factor (No. of servers)}}}
& & \multicolumn{3}{c|}{\textbf{IRN}} & \multicolumn{3}{c|}{\textbf{IRN + Timely}} & \multicolumn{3}{c|}{\textbf{IRN + DCQCN}} \\
\cline{3-11} 
& & Avg Slowdown & Avg FCT & 99\%ile FCT & Avg Slowdown & Avg FCT & 99\%ile FCT & Avg Slowdown & Avg FCT & 99\%ile FCT \\
\hline

\multirow{3}{*}{\emph{6 (54 servers)}} 
& IRN
&  8.24 & 0.0009 & 0.0153
&  4.73 & 0.0012 & 0.0185
&  5.19 & 0.0010 & 0.0207
\\
& $\frac{\text{IRN}}{\text{IRN+PFC}}$ 
& 0.513 & 0.640 & 0.612
& 0.995 & 0.976 & 0.968
& 1.009 & 1.005 & 0.966
\\
& $\frac{\text{IRN}}{\text{RoCE+PFC}}$ 
& 0.269 & 0.350 & 0.301
& 0.626 & 0.625 & 0.594
& 0.484 & 0.509 & 0.453
\\
\hline
\multirow{3}{*}{8 (128 servers)} 
& IRN
& 8.93 & 0.0011 & 0.0166 
& 4.98 & 0.0013 & 0.0195
& 5.36 & 0.0011 & 0.0234
\\
& $\frac{\text{IRN}}{\text{IRN+PFC}}$ 
& 0.497 & 0.601 & 0.515
& 1.000 & 0.993 & 0.985
& 1.010 & 0.998 & 0.992
\\
& $\frac{\text{IRN}}{\text{RoCE+PFC}}$ 
& 0.250 & 0.335 & 0.292
& 0.613 & 0.642 & 0.609
& 0.479 & 0.503 & 0.481
\\
\hline
\multirow{3}{*}{10 (250 servers)} 
& IRN
&  8.28 & 0.0010 & 0.0149  
&  4.48 & 0.0012 & 0.0177
&  4.87 & 0.0010 & 0.0211
\\
& $\frac{\text{IRN}}{\text{IRN+PFC}}$ 
& 0.486 & 0.601 & 0.547 
& 1.000 & 0.996 & 0.994 
& 1.015 & 1.010 & 1.012
\\
& $\frac{\text{IRN}}{\text{RoCE+PFC}}$ 
& 0.258 & 0.322 & 0.272  
& 0.651 & 0.664 & 0.631
& 0.477 & 0.491 & 0.445
\\
\hline

\end{tabular}
\caption{Robustness of IRN with varying fat-tree topology size (in terms of scale out factor or arity).}
\vspace{-5pt}
\label{tab:topoSize-details}
\end{table*}

\begin{table*}[t]
\centering
\scriptsize
\renewcommand{\arraystretch}{1.5}
\newcolumntype{P}[1]{>{\centering\arraybackslash}p{#1}}
\begin{tabular}[b]{|P{1.5cm}|P{1.5cm}|P{0.7cm}P{0.7cm}P{0.7cm}|P{0.7cm}P{0.7cm}P{0.7cm}|P{0.7cm}P{0.7cm}P{0.7cm}|}
\hline
\multirow{2}{*}{\parbox{1.5cm}{\centering \textbf{Workload pattern}}}
& & \multicolumn{3}{c|}{\textbf{IRN}} & \multicolumn{3}{c|}{\textbf{IRN + Timely}} & \multicolumn{3}{c|}{\textbf{IRN + DCQCN}} \\
\cline{3-11} 
& & Avg Slowdown & Avg FCT & 99\%ile FCT & Avg Slowdown & Avg FCT & 99\%ile FCT & Avg Slowdown & Avg FCT & 99\%ile FCT \\
\hline

\multirow{3}{*}{\parbox{1.5cm}{\centering \emph{Heavy-tailed (32B-3MB)}}}
& IRN
&  8.24 & 0.0009 & 0.0153
&  4.73 & 0.0012 & 0.0185
&  5.19 & 0.0010 & 0.0207
\\
& $\frac{\text{IRN}}{\text{IRN+PFC}}$ 
& 0.513 & 0.640 & 0.612
& 0.995 & 0.976 & 0.968
& 1.009 & 1.005 & 0.966
\\
& $\frac{\text{IRN}}{\text{RoCE+PFC}}$ 
& 0.269 & 0.350 & 0.301
& 0.626 & 0.625 & 0.594
& 0.484 & 0.509 & 0.453
\\
\hline
\multirow{3}{*}{\parbox{1.5cm}{\centering Uniform (500KB-5MB)}}
& IRN
& 18.93 & 0.0116 & 0.0428
& 18.91 & 0.0123 & 0.0337
& 16.06 & 0.0109 & 0.0557
\\
& $\frac{\text{IRN}}{\text{IRN+PFC}}$ 
& 0.313 & 0.334 & 0.170
& 0.955 & 0.957 & 0.919
& 0.993 & 0.988 & 0.974
\\
& $\frac{\text{IRN}}{\text{RoCE+PFC}}$ 
& 0.213 & 0.231 & 0.156
& 0.584 & 0.576 & 0.496
& 0.600 & 0.553 & 0.647
\\

\hline

\end{tabular}
\caption{Robustness of IRN with varying workload pattern.}
\vspace{-5pt}
\label{tab:workload-details}
\end{table*}

\subsection{Varying workload.} Our default workload comprised of a heavy-tailed mix of short messages (\eg for key-value lookups) and large messages (for storage or background applications). We also experimented with another workload pattern, comprising of medium to large sized flows with a uniform distribution, representing a scenario where RDMA is used only for storage or background tasks. Table~\ref{tab:workload-details} shows the results. We find that our key trends hold for this workload as well. 

Even when considering individual flow sizes in the range captured by our default workload, we did not observe any significant deviation from the key trends produced by the aggregated metrics.

\begin{table*}[t]
\centering
\scriptsize
\renewcommand{\arraystretch}{1.5}
\newcolumntype{P}[1]{>{\centering\arraybackslash}p{#1}}
\begin{tabular}[b]{|P{1.5cm}|P{1.5cm}|P{0.7cm}P{0.7cm}P{0.7cm}|P{0.7cm}P{0.7cm}P{0.7cm}|P{0.7cm}P{0.7cm}P{0.7cm}|}
\hline
\multirow{2}{*}{\parbox{1.5cm}{\centering \textbf{Buffer Size}}}
& & \multicolumn{3}{c|}{\textbf{IRN}} & \multicolumn{3}{c|}{\textbf{IRN + Timely}} & \multicolumn{3}{c|}{\textbf{IRN + DCQCN}} \\
\cline{3-11} 
& & Avg Slowdown & Avg FCT & 99\%ile FCT & Avg Slowdown & Avg FCT & 99\%ile FCT & Avg Slowdown & Avg FCT & 99\%ile FCT \\
\hline
\multirow{3}{*}{60KB} 
& IRN
& 9.58 & 0.0014 & 0.0225
& 5.75 & 0.0013 & 0.0213
& 6.81 & 0.0022 & 0.0464
\\
& $\frac{\text{IRN}}{\text{IRN+PFC}}$ 
& 0.285 & 0.371 & 0.351
& 0.723 & 0.597 & 0.596
& 0.848 & 0.829 & 0.883
\\
& $\frac{\text{IRN}}{\text{RoCE+PFC}}$ 
& 0.354 & 0.454 & 0.395
& 0.680 & 0.565 & 0.579
& 0.821 & 0.813 & 0.876
\\
\hline
\multirow{3}{*}{120KB} 
& IRN
& 8.87 & 0.0012 & 0.0191
& 4.99 & 0.0012 & 0.0192
& 5.68 & 0.0014 & 0.0324
\\
& $\frac{\text{IRN}}{\text{IRN+PFC}}$ 
& 0.343 & 0.410 & 0.340
& 0.863 & 0.821 & 0.794
& 0.951 & 0.945 & 0.932
\\
& $\frac{\text{IRN}}{\text{RoCE+PFC}}$ 
& 0.320 & 0.411 & 0.353
&  0.603 & 0.578 & 0.562
& 0.689 & 0.697 & 0.692
\\
\hline
\multirow{3}{*}{\emph{240KB}} 
& IRN
&  8.24 & 0.0009 & 0.0153
&  4.73 & 0.0012 & 0.0185
&  5.19 & 0.0010 & 0.0207
\\
& $\frac{\text{IRN}}{\text{IRN+PFC}}$ 
& 0.513 & 0.640 & 0.612
& 0.995 & 0.976 & 0.968
& 1.009 & 1.005 & 0.966
\\
& $\frac{\text{IRN}}{\text{RoCE+PFC}}$ 
& 0.269 & 0.350 & 0.301
& 0.626 & 0.625 & 0.594
& 0.484 & 0.509 & 0.453
\\
\hline
\multirow{3}{*}{480KB} 
& IRN
& 8.63 & 0.0008 & 0.0127
& 4.66 & 0.0012 & 0.0185
& 4.99 & 0.0010 & 0.0206
\\
& $\frac{\text{IRN}}{\text{IRN+PFC}}$ 
& 0.853 & 0.953 & 0.945
& 1.005 & 1.004 & 1.004
& 1.003 & 1.001 & 0.955
\\
& $\frac{\text{IRN}}{\text{RoCE+PFC}}$ 
& 0.223 & 0.315 & 0.285
& 0.621 & 0.657 & 0.614
&0.310 & 0.442 & 0.405
\\
\hline

\end{tabular}
\caption{Robustness of IRN with varying per-port buffer size}
\vspace{-5pt}
\label{tab:bufferSize-details}
\end{table*}

\subsection{Varying buffer size.} Table~\ref{tab:bufferSize-details} shows the robustness of our basic results as the buffer size is varied from 60KB to 480KB. We find that as the buffer size is decreased, the drawbacks of using PFC increases, due to more pauses and greater impact of congestion spreading.~\footnote{Decreasing just the PFC threshold below its default value also has a similar effect.}  In general, as the buffer size is increased, the difference between PFC-enabled and PFC-disabled performance with IRN reduces (due to fewer PFC frames and packet drops), while the benefits of using IRN over RoCE+PFC increases (because of greater relative reduction in queuing delay with IRN due to BDP-FC). We expect to see similar behaviour in shared buffer switches.

\begin{table*}[t]
\centering
\scriptsize
\renewcommand{\arraystretch}{1.5}
\newcolumntype{P}[1]{>{\centering\arraybackslash}p{#1}}
\begin{tabular}[b]{|P{1.5cm}|P{1.5cm}|P{0.7cm}P{0.7cm}P{0.7cm}|P{0.7cm}P{0.7cm}P{0.7cm}|P{0.7cm}P{0.7cm}P{0.7cm}|}
\hline
\multirow{2}{*}{\parbox{1.5cm}{\centering $RTO_{high}$}}
& & \multicolumn{3}{c|}{\textbf{IRN}} & \multicolumn{3}{c|}{\textbf{IRN + Timely}} & \multicolumn{3}{c|}{\textbf{IRN + DCQCN}} \\
\cline{3-11} 
& & Avg Slowdown & Avg FCT & 99\%ile FCT & Avg Slowdown & Avg FCT & 99\%ile FCT & Avg Slowdown & Avg FCT & 99\%ile FCT \\
\hline


\multirow{3}{*}{\emph{320$\mu$s}} 
& IRN
&  8.24 & 0.0009 & 0.0153
&  4.73 & 0.0012 & 0.0185
&  5.19 & 0.0010 & 0.0207
\\
& $\frac{\text{IRN}}{\text{IRN+PFC}}$ 
& 0.513 & 0.640 & 0.612
& 0.995 & 0.976 & 0.968
& 1.009 & 1.005 & 0.966
\\
& $\frac{\text{IRN}}{\text{RoCE+PFC}}$ 
& 0.269 & 0.350 & 0.301
& 0.626 & 0.625 & 0.594
& 0.484 & 0.509 & 0.453
\\
\hline

\multirow{3}{*}{640$\mu$s} 
& IRN
& 8.38 & 0.0010 & 0.0162
& 4.77 & 0.0012 & 0.0188
& 5.23 & 0.0010 & 0.0206
\\
& $\frac{\text{IRN}}{\text{IRN+PFC}}$ 
&  0.522 & 0.675 & 0.649
&  1.003 & 0.987 & 0.987
&  1.018 & 0.995 & 0.964

\\
& $\frac{\text{IRN}}{\text{RoCE+PFC}}$ 
& 0.274 & 0.369 & 0.319
& 0.631 & 0.631 & 0.605
& 0.488 & 0.503 & 0.452

\\
\hline

\multirow{3}{*}{1280$\mu$s} 
& IRN
& 8.74 & 0.0011 & 0.0194
& 4.79 & 0.0012 & 0.0194
& 5.24 & 0.0010 & 0.0207
\\
& $\frac{\text{IRN}}{\text{IRN+PFC}}$ 
& 0.544 & 0.779 & 0.776
&  1.008 & 1.007 & 1.016
& 1.020 & 1.007 & 0.968
\\
& $\frac{\text{IRN}}{\text{RoCE+PFC}}$ 
&  0.285 & 0.426 & 0.382
& 0.634 & 0.644 & 0.623
& 0.489 & 0.510 & 0.453
\\
\hline

\end{tabular}
\caption{Robustness of IRN to higher $RTO_{high}$ value.}
\vspace{-5pt}
\label{tab:timeout-details}
\end{table*}

\begin{table*}[t]
\centering
\scriptsize
\renewcommand{\arraystretch}{1.5}
\newcolumntype{P}[1]{>{\centering\arraybackslash}p{#1}}
\begin{tabular}[b]{|P{1.5cm}|P{1.5cm}|P{0.7cm}P{0.7cm}P{0.7cm}|P{0.7cm}P{0.7cm}P{0.7cm}|P{0.7cm}P{0.7cm}P{0.7cm}|}
\hline
\multirow{2}{*}{\parbox{1.5cm}{\centering $N$ for using $RTO_{low}$}}
& & \multicolumn{3}{c|}{\textbf{IRN}} & \multicolumn{3}{c|}{\textbf{IRN + Timely}} & \multicolumn{3}{c|}{\textbf{IRN + DCQCN}} \\
\cline{3-11} 
& & Avg Slowdown & Avg FCT & 99\%ile FCT & Avg Slowdown & Avg FCT & 99\%ile FCT & Avg Slowdown & Avg FCT & 99\%ile FCT \\
\hline

\multirow{3}{*}{\emph{3}} 
& IRN
&  8.24 & 0.0009 & 0.0153
&  4.73 & 0.0012 & 0.0185
&  5.19 & 0.0010 & 0.0207
\\
& $\frac{\text{IRN}}{\text{IRN+PFC}}$ 
& 0.513 & 0.640 & 0.612
& 0.995 & 0.976 & 0.968
& 1.009 & 1.005 & 0.966
\\
& $\frac{\text{IRN}}{\text{RoCE+PFC}}$ 
& 0.269 & 0.350 & 0.301
& 0.626 & 0.625 & 0.594
& 0.484 & 0.509 & 0.453
\\
\hline

\multirow{3}{*}{10} 
& IRN
& 8.26 & 0.0010 & 0.0157
& 4.75 & 0.0012 & 0.0187
& 5.13 & 0.0010 & 0.0208
\\
& $\frac{\text{IRN}}{\text{IRN+PFC}}$ 
&  0.515 & 0.651 & 0.629
&  0.999 & 0.983 & 0.979
&  0.997 & 0.989 & 0.974

\\
& $\frac{\text{IRN}}{\text{RoCE+PFC}}$ 
& 0.270 & 0.356 & 0.310
& 0.628 & 0.629 & 0.600
& 0.478 & 0.500 & 0.456

\\
\hline

\multirow{3}{*}{15} 
& IRN
& 8.25 & 0.0010 & 0.0154
& 4.72 & 0.0012 & 0.0187
& 5.22 & 0.0010 & 0.0218
\\
& $\frac{\text{IRN}}{\text{IRN+PFC}}$ 
& 0.514 & 0.653 & 0.616
& 0.992 & 0.981 & 0.981
& 1.015 & 1.019 & 1.019
\\
& $\frac{\text{IRN}}{\text{RoCE+PFC}}$ 
& 0.270 & 0.357 & 0.303
& 0.624 & 0.628 & 0.602
& 0.487 & 0.516 & 0.477
\\
\hline

\end{tabular}
\caption{Robustness of IRN to higher $N$ value for using $RTO_{low}$.}
\vspace{-5pt}
\label{tab:varyN-details}
\end{table*}

\subsection{Varying parameters.}
We finally present the robustness of our results as we vary some of IRN's parameters from their default values. In particular, Table~\ref{tab:timeout-details} captures the effect of over-estimating the $RTO_{high}$ value to 2$\times$ and 4$\times$ the ideal, and Table~\ref{tab:varyN-details} shows results with higher $N$ values for using $RTO_{low}$. We find that changing these parameters produces very small differences over our default case results, showing that IRN is fairly robust to how its parameters are set. 

\section{Additional Implementation Details}

\subsection{Conditions for Write Completion}

The RDMA specification (Sec o9-20~\cite{ibaspec}) clearly states that an application shall not depend on the contents of an RDMA Write buffer at the responder, until one of the following has occurred: (1) arrival and completion of the last RDMA Write request packet when used with Immediate data; (2) arrival and completion of a subsequent Send message; (3) update of a memory element by a subsequent Atomic operation. IRN design guarantees that meeting any of these conditions would automatically imply that previous Writes have completed. As discussed before, it supports Write with immediate, where a CQE for the request is not released to the responder's application until all packets up until the last packet of the request have been received. Likewise, the CQE for a subsequent send will not be released to the application until all previous packets have arrived. The Atomic request packet will wait in the Read/Atomic WQE buffer without being processed until all previous packets have arrived.  
\subsection{IRN Support for Shared Receive Queues}
\label{app:srq}

\paragraphb{For Send/Receive:} As mentioned previously, the QP maintains a running total of allotted \emph{recv\_WQE\_SN}. But rather than allotting it as soon as a new receive WQE is posted as described before, with SRQ, we allot it when new recv WQEs are dequeued from SRQ. Suppose we start with \emph{recv\_WQE\_SN} of 0. The first send packet arrives in order with \emph{recv\_WQE\_SN} of 0. We will dequeue one WQE from SRQ, allot it \emph{recv\_WQE\_SN} of 0 and increment the \emph{recv\_WQE\_SN} state maintained by the QP. Suppose after this some intermediate send requests are lost and we see next send packet with \emph{recv\_WQE\_SN} of 4. We will then dequeue 4 Receive WQEs from SRQ, allot them \emph{recv\_WQE\_SN} of 1,2,3 and 4 and will use the 4th Receive WQE to process the packet. 

\paragraphb{For Write with Immediate:} Write with immediate do not require the Receive WQE to process incoming packets. The Receive WQE just needs to expire when the entire message is received. If there are no outstanding Sends (and already dequeued Receive WQEs  waiting at the QP), then dequeue the first available WQE from SRQ and expire it to generate a completion event, when all packets of the Write with Immediate request have been received. If there are outstanding Receive WQEs at the Responder QP, expire the first Receive WQE that is outstanding (the fact that we check for message completion using our bitmap \emph{in order} guarantees that the first Receive WQE is the correct WQE that needs to be expired).

\subsection{IRN Support for End-to-End Credits}
\label{app:e2ecredits}

\paragraphb{With current RoCE NICs:} For messages that need a Receive WQE, an end-to-end credit scheme is used, where the acks piggy-back the information about the number of Receive WQEs (or credits) remaining. When the responder runs out of credits, it can still send the first packet of a Send message or all packets of a Write with Immediate message as a probe. If the receiver has new WQEs the operation executes successfully and the new credit is communicated via the acknowledgement packet. Otherwise, an RNR (receiver not ready) NACK is sent which results in go-back-N. 

\paragraphb{With IRN:} This can be done with IRN too. Although when an out-of-sequence probe packet is received without credits (with no Receive WQE), it should be dropped at the receiver. For example, if there is only one Receive WQE at the responder and the requester sends two Send messages (the first as a valid message, the second as a probe). If the first message is lost, the second message should be dropped instead of placing it in first message's memory address (which would be wrong to do) or sending an RNR NACK (which would be ill-timed). The first message will be sent again due to loss recovery and things will get back on track.

\subsection{NACKs due to Other Errors}
\label{app:other-errors}

This is a generalization of the case described in Appendix~\ref{app:e2ecredits}. Current RoCE NICs generate a NACK for out-of-sequence packets, the requester treats them as errors and does a go-back-n on receiving them. With IRN, we consider out-of-sequence NACKs  as normal behaviour and treat them differently (as described in \S\ref{sec:irnTransportDesign}). But NACKs can still be generated for other reasons such as ``receiver not ready''. IRN will do a go-back-N on receiving such a NACK. If an out-of-sequence packet will result in generation of such an error NACK at the responder, IRN will discard that packet at the responder, without processing it and without sending a NACK.

\subsection{Supporting Send with Invalidate}

 This operation is used to invalidate the use of a remote memory region. If this packet arrives and is executed before previous Writes on the invalidated region, the Write operation would die. To avoid this, with IRN can enforce a fence before the Send with Invalidate operations.
\end{appendices}

\footnotesize\bibliographystyle{plain} 
\bibliography{main}

\balance

\end{document}